\documentclass[lettersize,journal]{IEEEtran}
\usepackage{cite}
\usepackage{amsmath,amssymb,amsfonts}
\usepackage{algorithmic}
\usepackage{graphicx}
\usepackage{textcomp}
\usepackage{booktabs}
\usepackage{array}
\usepackage{makecell}
\usepackage{multirow}
\usepackage{graphicx}
\usepackage{color,soul}

\begin{document}


\title{Integrating Counter-UAS Systems into the UTM System for Reliable Decision Making}
\author{Abdulhadi Shoufan and Ruba Alkadi
\thanks{ Center of Cyber-Physical Systems, Khalifa University, Abu Dhabi, UAE (emails: {Abdulhadi.shoufan, Ruba.alkadi  }@ku.ac.ae)}}

\markboth{This work is submitted to an IEEE journal. }%
{Shell \MakeLowercase{\textit{et al.}}: A Sample Article Using IEEEtran.cls for IEEE Journals}


\maketitle

\begin{abstract}
Despite significant progress, the deployment of UAV technology in commercial and civil applications is still lagging. This is essentially due to the risks associated with drone flights and the lack of coordinated technologies that would mitigate these risks. While Unmanned Aircraft System Traffic Management systems (UTM) are being developed worldwide to enable safe operation, the counter-drone technology operates on an all-enemy basis and regards any sighted drone as a threat. This situation is essentially caused by the lack of information exchange between stakeholders. Without the exchange of relevant information, a counter-drone system can misclassify drones and initiate erroneous interdiction procedures.
This paper proposes a system that integrates counter-drone technology into the UTM system for information exchange and coordination using a set of clarification protocols towards accountable response to sighted drones. The system functionality and performance were evaluated by simulation.

\end{abstract}


\section{Introduction} \label{sec:introduction}

The drone market is growing rapidly with diverse applications in construction, agriculture, insurance, oil and gas industry, film making, sky photography, parcel delivery, journalism, security, law enforcement, and civil defense{\cite{hassanalian2017classifications}}. Despite this, the operation of UAVs in urban areas is still in the exploratory stage \cite{xu2020recent}. That is, we neither get our online orders delivered by UAVs nor can we ride a taxi drone, although the technical capabilities of today's UAVs allow for the deployment in such applications \cite{courtin2018feasibility}. 
Indeed, flying a drone is associated with security, privacy, and safety risks which have limited the progress and penetration of UAV applications across many sectors in the urban airspace  \cite{altawy2016security, lin2018security}. Safety is, without doubt, the most urgent requirement when it comes to drone operation. This is confirmed by the frequent reports on drone incidents and intrusions worldwide \cite{lykou2020defending, dedrone-list-of-incidents, Gatwick}.

\begin{figure}
    \centering
    \includegraphics[width=8 cm ]{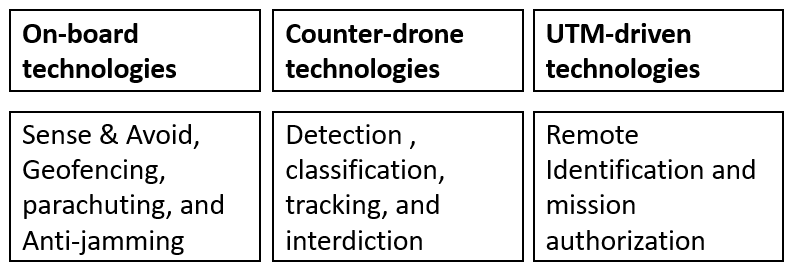}
    \caption{Three technology categories for UAV and airspace safety}
    \label{fig-safety-technologies}
\end{figure}

To support safe drone operations and counter violations many technologies are available. These technologies can be divided broadly into three categories as depicted in Fig.~\ref{fig-safety-technologies}. The first category encompasses preventive onboard solutions such as sense\&avoid and geofencing \cite{dentler2016real}, as well as response technologies such as parachuting and self-recovery systems \cite{kim2017self}. The second category addresses illegal UAVs and includes counter-drone systems from detection to interdiction. Detection and classification use different modalities such as radar, images, acoustic, and radio-frequency signals \cite{anwar2019machine, taha2019machine}. Malicious drones can be deactivated using diverse interdiction technologies such as jamming, catching, or shooting \cite{multerer2017low, rothe2019concept}. Finally, UTM-driven technologies, including mission authorization systems and remote identification, have been put forward recently to coordinate UAV activities in the low-altitude airspace \cite{Unmanned2019NASA, barrado2020u}. 

While these technologies are supposed to complement each other towards safe airspace, the current implementations of category-2 and category-3 technologies lack the necessary level of integration and appear to follow conflicting goals: While UTM-driven technologies promote the legalization of drone operations, the counter-drone systems take a skeptical position and consider any drone in the sky as a threat. This suggests that information exchange between UTM systems and counter-UAS systems is mandatory to avoid conflicts. These observations were already reported in the recent survey of counter-drone technology \cite{park2021survey}. Hence, the first contribution of this paper is to propose a novel architecture for integrating CUAS systems in the UTM system (Section \ref{sec:system-architecture}). This architecture will allow the CUAS system to obtain information about the registration status of the sighted drone and whether it is authorized to perform the observed flight. With this information, the CUAS system can classify the drone as cooperative and tolerate its existence in the operation zone of the CUAS system. To access relevant information in the UTM databases, a CUAS system requires an initial knowledge about the drone. We propose to obtain this initial knowledge from the drone itself through the broadcast remote identification. Remote identification (RID) is becoming a core technology of UTM systems in many countries.


No matter how reliable these technologies are, technical and non-technical errors can occur: 
\begin{enumerate}
    \item The RID module on board the drone can fail.
    \item The RID receiver of the CUAS system can fail.
    \item A UTM system operator can fail to update the  databases or to do so on time.
    \item An inquiry to the UTM databases can fail for technical reasons.
    \item The drone operator can exceed the permitted flight time by mistake or for an urgent reason.
    \item The drone operator can deviate from the approved mission trajectory by mistake or for mandatory reasons.
\end{enumerate}

In none of these cases should a CUAS system immediately classify a drone as uncooperative and start an interdiction operation. On the other hand, the outcomes of such cases are indeed ambiguous and can indicate uncooperative or malicious drone operations that should be countered. These thoughts suggest that a counter-drone system needs more information to clarify such cases and classify them correctly towards a responsible decision. Therefore, the second contribution of this paper is to present a post-detection model for the drone and a set of protocols that help in clarifying ambiguous cases, see Section \ref{sec:drone-model}.  

The developed models and protocols are evaluated by simulation. First, we used an appropriate Simulink toolkit to test the correctness of the model and protocols. Then, we implemented the same using Node.js to assess the performance of the system in a networked mode. The simulation setup and results are present in Section \ref{sec-simulation}.

\section{Related Work} 

\subsection{UTM Systems}

The concept of unmanned traffic management (UTM) refers to an ecosystem for controlling the operation of unmanned aerial systems \cite{Unmanned2019NASA, xu2020recent}. Data exchange is at the core of UTM where authorized unmanned service suppliers (USS) provide cloud-based services to the different stakeholder. Examples of these services include UAV control \cite{ zhou2021control}, efficient and fair unmanned traffic control \cite{ chin2021efficiency}, flight planning and scheduling \cite{liu2020iterative}, geofencing \cite{stevens2020geofence}, path optimization and collision avoidance \cite{chakrabarty2019vehicle}, weather and contingency management \cite{reiche2021initial, neogi2021assuring} , orchestrating of UAV services \cite{bekkouche2019toward}, and supporting the internet-of-drones (IoD) \cite{allouch2021utm}.

UTM systems are still in the development stage. Some authors have highlighted several challenges and issues in the design of these systems. Wolter et al. pointed out multiple obstacles in the current experimental setups which relate to standardization, information quality, and the transition from human-centric design to automation \cite{wolter2020human}. Several authors addressed the security of UTM systems and presented their vulnerabilities to various cyber and physical attacks \cite{allouch2021utm}. The lack of a legal framework for UTM system operations was highlighted in \cite { ryan2020legal }. The authors described the fundamentals of such a legal framework which should provide the needed certainty for all stakeholders.

\subsection{Counter-Drone Systems}

The counter-drone industry has boomed in recent years. A report published by the "Center for the Study of the Drone" at Bard College shows that there are 537 counter-unmanned aerial systems (CUAS) on the market \cite{counter-drone-technologies-survey}. Researchers showed wide interest in this field, especially regarding the detection and classification of small UAVs. Different technologies were proposed including radar, radio-frequency detection, acoustic systems, and computer vision \cite{drozdowicz201635, nguyen2016investigating, anwar2019machine, aker2017using}.
Parallel to the advances in detection and classification technologies, some researchers investigated technical solutions for drone interdiction. Wyder et al. classified these technologies according to their impact on the target drone into three main categories: signal interception, propeller restriction, and aerial takedown \cite{wyder2019autonomous}.
Due to its undisruptive nature, signal interception has received substantial attention for interdiction in urban areas. Depending on the operation mode of the drone, Roth et al. identified two methods of signal interception-based interdiction: drone hacking and GPS spoofing \cite{rothe2019concept}. Propeller restriction refers to capturing uncooperative drones usually using a net. The net could be launched either manually by a skilled operator on the ground or autonomously by another flying drone \cite{ armstrong2019interdiction}. Finally, a variety of aerial takedown technologies was presented. These include hunting by eagles \cite{ o2019no } and shooting by machine guns or laser \cite{ rothe2019concept }. 

\subsection{Coordination between UTM and Counter-Drone Systems}

Despite the strong relationship between UTM and counter-drone systems, the interaction between these systems remained unaddressed in the literature. Recently, Park et. al \cite{park2021survey} surveyed more than 180 publications on the literature of counter-drone systems. In their concluding remarks, they highlighted the necessity of integrating a well-defined drone identification scheme into counter drone systems. Sandor has highlighted that with the spread of UTM systems, we need to define the problems, the scope, and the operational environment \cite{sandor2019challenges}. The author defined and classified many functions related to UTM and interestingly suggested surveillance as a UTM service with different technologies for the detection of cooperative and non-cooperative vehicles. However, the author did not take into consideration the critical function of a counter-drone system which is interdiction. Apart from this, we are not aware of any literature that addressed the coordination of the UTM and counter-drone systems' activities.   

\section{System Architecture}

\label{sec:system-architecture}

\begin{figure}
	\centering
	\includegraphics[width=9 cm]{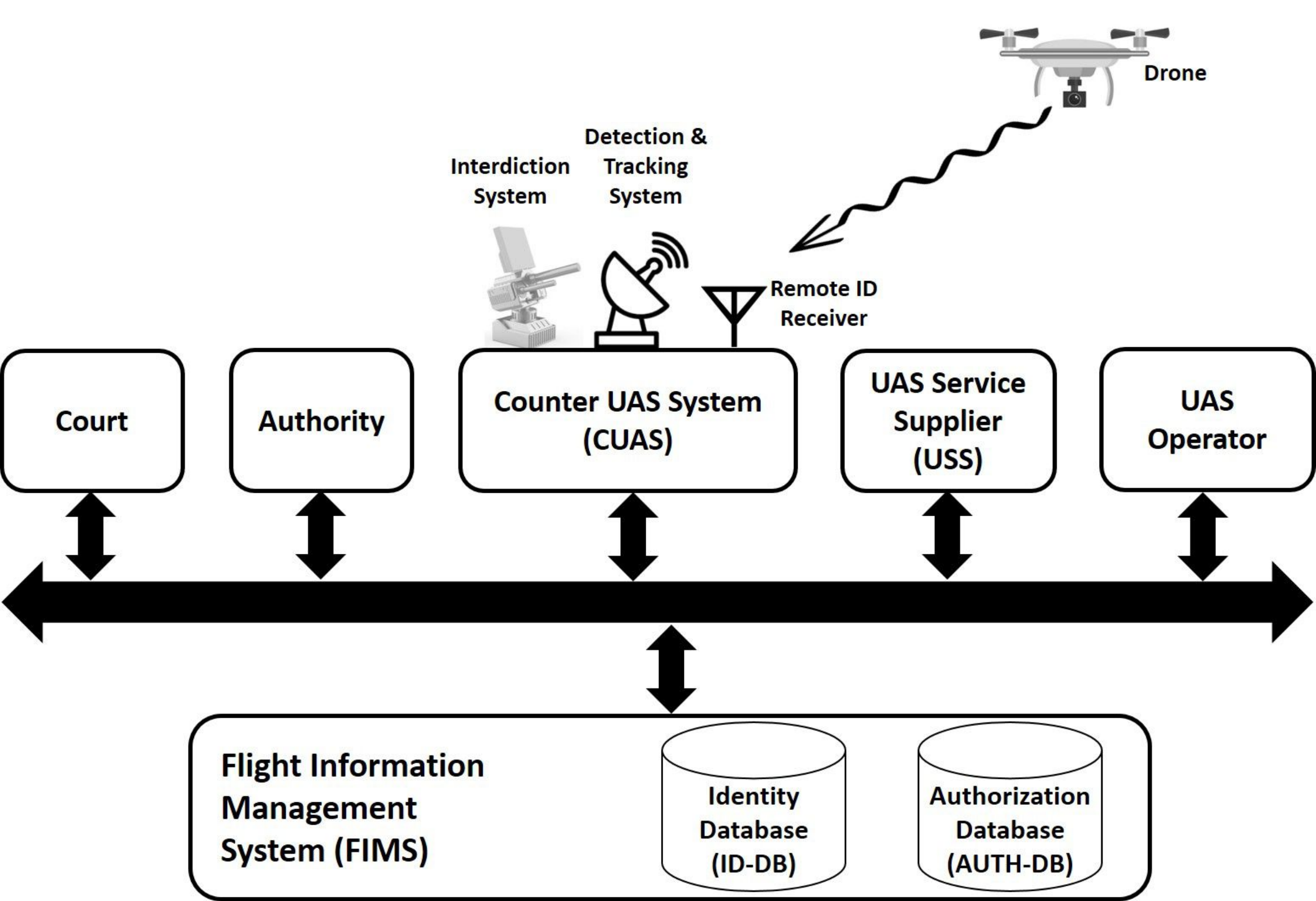}
	\caption{System Architecture for Coordinated Drone Interdiction}
	\vspace{-10pt}
	\label{fig-architecture}
\end{figure}

The proposed system architecture relies on and extends the UAS traffic management system (UTM) released by NASA \cite{Unmanned2019NASA}. The proposed architecture in Fig. \ref{fig-architecture} aims to grant CUAS systems access to the Flight Information Management System (FIMS). The latter should contain at least two databases: a database of the identities of registered drones (ID-DB) and a database of authorized missions (AUTH-DB). In the following, we first describe the system agents in brief. Then, we describe the databases and how the CUAS system interacts within the UTM system.

The \textit{Authority} in our architecture refers to the agent that regulates and oversees civil airspace to assure safe and secure operation by providing air navigation services and facilitating air connectivity. All drones should be registered at the authority. Additionally, operators can obtain mission authorization directly from the authority or from authorized service suppliers. The authority along with such service suppliers keeps relevant information about drone registrations and mission authorizations in the ID-DB and AUTH-DB databases.

The \textit{Unmanned Service Supplier} (USS) is an entity that provides services to subscribed UAS operators to help them meet the operational requirements specified by the authority. Operation planning, strategic and tactical de-confliction, remote identification, and airspace authorization are examples of the services provided by a USS. An operator may subscribe to one or more USS's to use multiple services.

The \textit{UAS operator} is a primary agent in the UTM system who is responsible for operating the drone in different modes for various purposes. UAS operators may opt to utilize a USS or to provide their own services to meet regulatory obligations. So, it is the operator's responsibility to meet the requirements established for the type of operation and the associated airspace volume or route.

The \textit{Counter Unmanned Aerial System} (CUAS) is an active agent in the UTM architecture according to our proposal. By far, the deployment of anti-drone technology against civil drones has been served as individual efforts by different agents interested in protecting their territories against malicious drones or by law enforcement for protecting public safety. 
We advocate for integrating the CUAS technology into the UTM system to enable the exchange of information between relevant agents and support coordinated efforts for cooperative vs. uncooperative drone identification.

Finally, the \textit{Court} is added as an agent to the system to benefit from the information infrastructure for evidence collection in case of legal conflicts. Basically, following to any interdiction process, the authority can file a lawsuit to the court to initiate forensics work. Similarly, a UAS operator might file a lawsuit in case her or his vehicle was intercepted, fined, or jammed unlawfully.

\subsection{Identity Database (ID-DB)} 
\label{sec-ID-DB}

Aviation authorities worldwide are mandating the deployment of remote identification to promote safe and accountable drone operations. The European Union Aviation Safety Agency (EASA) has published related regulations in March 2019 (Commission delegated regulation (EU) 2019/945) that were amended in April 2020 ((EU) 2020/1058). These regulations mandate that all unmanned aircraft should be equipped with a remote identification system~\cite{eu1058}. The FAA in the USA has published a final rule for remote identification in January 2021 \cite{faa-final-rule}. According to this, the remote identification message should include information about the drone's identity, location, altitude, and velocity in addition to the control station location and elevation, a time mark, and emergency status. A database with Remote ID information should provide three levels of access: 
\begin{enumerate}
	\item Level 1: Information available to the public (the UAS unique identifier).
	\item Level 2: Information available to designated public safety and airspace management officials (personally identifiable information).
	\item Level 3: Information available to the FAA and certain Federal, State, and local agencies (tracking data).	
\end{enumerate}

In our architecture, we propose including this database in the FIMS. Certified CUAS systems should be given access to this database on level 2 or 3 depending on the predetermined authority of these systems. We further assume that this database is available and its content is protected from security attacks. Specifically, Remote ID entries in the database are assumed to be authentic. This means that a CUAS can use the ID-DB to verify the authenticity of a received Remote ID. 

\subsection{Authorization Database (AUTH-DB)} 
\label{sec-AUTH-DB}

CUAS systems are usually used in protected areas such as controlled zones around airports. But even in such areas, drones can have a reason to fly, for example, to help firefighters nearby.  So, the CUAS needs a possibility to check whether a detected drone has permission to fly in this area and time. The FAA has implemented a system referred to as Low Altitude Authorization and Notification Capability (LAANC) \cite{laanc}. This system automates the application and approval process for airspace authorizations. Like Remote ID, the LAANC is considered a part of the UTM ecosystem. UAS service suppliers support the authorization process: A drone pilot submits a request through a LAANC USS. The request is checked against multiple airspace data sources by the FAA. If approved, the pilot receives the authorization in near real-time. 

For our purpose, we propose that all mission authorizations should be logged in AUTH-DB. An entry in this database should include information about the drone ID and the mission date, time, and area. Operators of counter-drone systems should have access to the AUTH-DB to verify the performance authorization of a detected drone. The AUTH-DB must be protected against security attacks such as entry manipulation. 

\subsection{CUAS Interactions within the UTM System} 
\label{sec-AUTH-DB}

The proposed architecture in Fig. \ref{fig-architecture} allows the CUAS to interact with different agents in the UTM system and to perform different checks using the identity and the authorization databases towards cooperative vs. uncooperative identification. For this, we make the following assumptions:   

\begin{enumerate}
	\item Any drone in flight should share its ID remotely.
	\item A CUAS has the necessary technology to receive the Remote ID and check its authenticity and validity by accessing the ID-DB. 
	\item The ID-DB should be kept up-to-date by the authority or any certified USS. Nonetheless, legacy/expired IDs are not removed from the database.
	\item Any drone flying in the operation zone of a CUAS (i.e., the zone in which the CUAS is permitted to interdict), is required to have a performance authorization.
	\item A performance authorization includes at least information about the drone ID as well as the date, time, and area of flight.
	\item The CUAS has access to the AUTH-DB to verify whether a detected drone is authorized to fly in the respective area and time. 
	\item The AUTH-DB should be kept up-to-date by the authority or any certified agent. 
\end{enumerate} 

In the following section, the CUAS interaction with the UTM system will be detailed by defining a post-detection model for the drone. 

\section{Post-Detection Model for Drones} \label{post-detection}
\label{sec:drone-model}

\begin{figure*}[hbtp]

\includegraphics[width=12 cm]{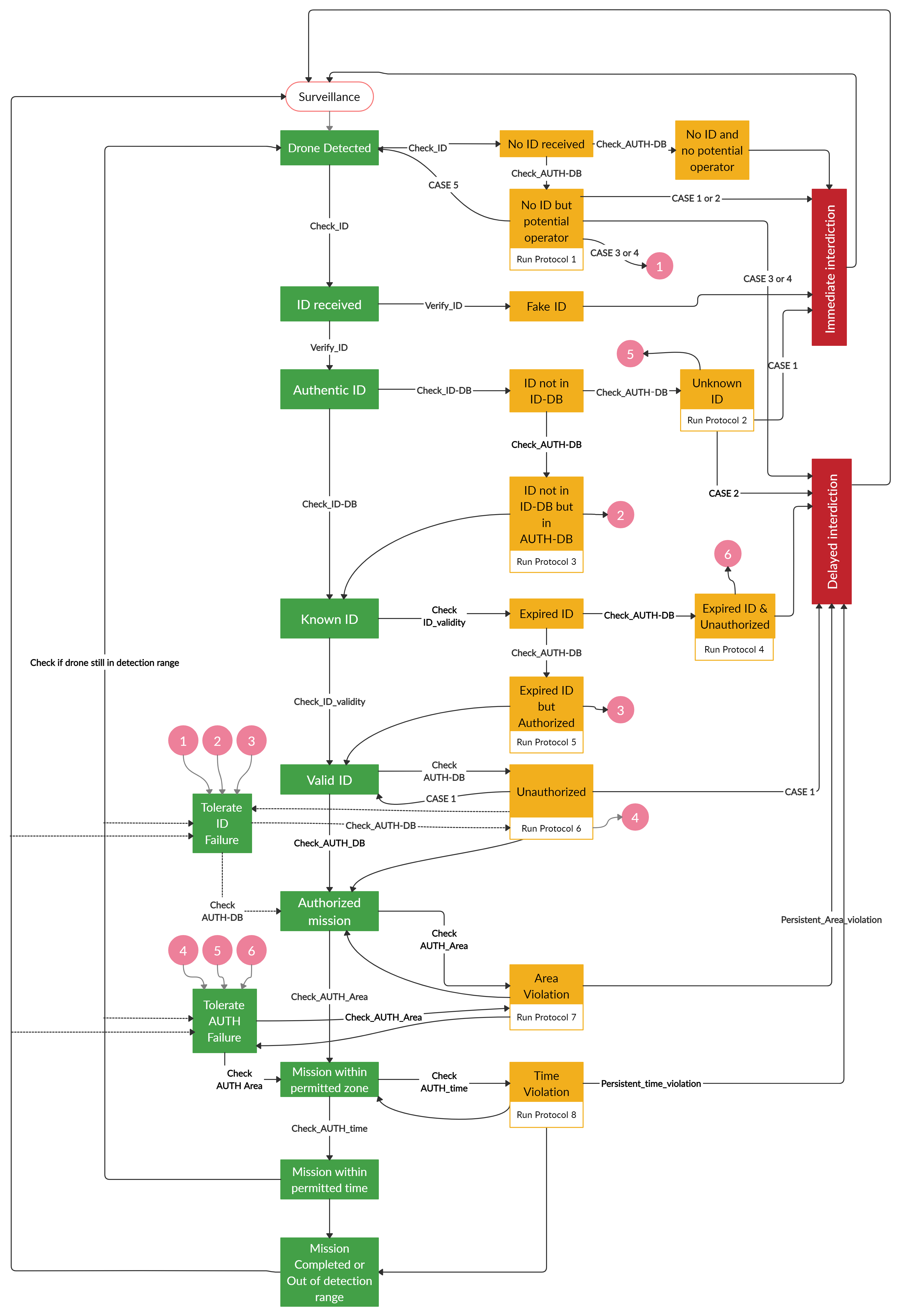}
\centering
	\caption{ {Post-detection model for drones from CUAS perspective. The drone is modeled as a finite-state machine. A cooperative drone passes multiple checks transiting from one green state to another. A drone that fails some check transits to an orange state that can trigger some clarification protocol.}}
	\label{fig-model}
\end{figure*}

Based on the architecture presented in the previous section, we model a drone from a CUAS perspective using the finite state machine illustrated in Fig. \ref{fig-model}.
The transition from one state to another is triggered either by the result of some checks that the CUAS performs independently or by the outcome of some protocol execution between the CUAS and other agents. In the beginning, the system is set in the \textit{Surveillance} state. When the CUAS identifies an object as a drone, it transits to the state \textit{Drone Detected}. The machine states are highlighted using three colors:

\begin{enumerate}
	\item The green color indicates the desired state which is reached through a successful check of some conditions. For example, when the CUAS verifies that the drone is transmitting a Remote ID, the next state would be \textit{ID Received}. Furthermore, the model has two special green states in which the drone is tolerated despite some failures, as will be described in Section \ref{sec:protocols}.
	\item A drone that reaches a red state is an uncooperative one. It will be subject to an approved interdiction, which can be started immediately or after a timeout.
	\item The orange color describes an ambiguous state of the detected drone. The CUAS essentially tries to clarify this state by executing some protocol involving other system agents. As a result, the drone either returns to a green state or moves to a red state.
	
\end{enumerate}

We first explain the case of a cooperative drone, which is represented by the vertical green path. When a drone is detected, the CUAS receives its Remote ID. The successful reception of the ID is followed by verifying its authenticity. If the ID is authentic, the CUAS checks if it is available in the identity database (ID-DB). If this is the case, the ID's expiry date is checked. If the ID is valid, the CUAS accesses the authorization database (AUTH-DB) to verify whether the drone is authorized to perform the observed mission. If the check is successful, the CUAS verifies if the drone is complying with the mission plan regarding the area and time. If this is the case, the FSM returns to the state \textit{Drone Detected}. The CUAS can repeat these checks until the drone has completed its mission or disappeared from the detection range.  

We now explain in brief what happens when any of the previous checks fails. Detailed descriptions of the different cases and protocols will follow in Section \ref{sec:protocols}. First, when the CUAS does not receive an ID from the drone (\textit{No ID Received}), the CUAS uses knowledge about the flight time and drone location to inquire about a potential operator in the authorization database (AUTH-DB).  If no such operator is found, the drone state changes to \textit{No ID \& No Potential Operator}. Depending on the CUAS policy and the authority, an \textit{Immediate Interdiction} process can be started. In contrast, if the CUAS finds a potential operator in the AUTH-DB, the machine transits to the state \textit{No ID But Potential Operator}. Herein, the CUAS reads the ID of the potential operator and initiates \textbf{Protocol 1} to verify whether this ID belongs to the actual operator who is flying the detected drone. Executing this protocol can lead to one of four results:
\begin{enumerate}
	\item Restoring the ID transmission by the operator.
	\item Tolerating the ID transmission failure.
	\item Immediate interdiction of the drone.
	\item Delayed interdiction of the drone.
\end{enumerate}

The CUAS should verify the authenticity of a received remote identification. When this verification is not successful, the CUAS classifies the drone as uncooperative. The next state is \textit{Fake ID} leading to the state \textit{Immediate Interdiction}. In contrast, if the ID is authentic, the CUAS looks for a related entry in the ID-DB. If no such ID is found, the next state will be \textit{ID not in ID-DB}. The CUAS tries to clarify this situation by checking if the unknown ID has a corresponding entry in the AUTH-DB. If this is not the case, the CUAS moves to the state \textit{Unknown ID} and executes \textbf{Protocol 2} to clarify the reason for the missing ID and the missing performance authorization. Executing this protocol can lead to one of four results:
\begin{enumerate}
	\item Restoring technical issues causing database misses.
	\item Tolerating the missing ID and authorization.
	\item Immediate interdiction of the drone.
	\item Delayed interdiction of the drone.
\end{enumerate}

In contrast, if AUTH-DB has a related entry, the CUAS initiates \textbf{Protocol 3} to clarify the reason for the missing ID in the ID-DB. Executing this protocol may lead to restoring the technical issue in the ID-DB or tolerating the missing ID.

An authentic and known ID can be invalid for different reasons. For example, the operator may miss renewing the registration (if required) or the authority/USS may miss updating the ID-DB on time. When the CUAS finds out that the received ID has expired (\textit{Expired ID}), it checks if the AUTH-DB has authorization for this ID in the current time and area. If no entry is found, the drone moves to the state \textit{Expired ID \& Unauthorized}. \textbf{ {Protocol 4}} is initiated to clarify this state. Three outcomes are possible:
\begin{enumerate}
	\item Restoring the technical issues in both databases.
	\item Tolerating the expired ID and the missing authorization.
	\item Delayed interdiction.
\end{enumerate}

In contrast, if the CUAS finds a related entry in the AUTH-DB, it initiates \textbf{Protocol 5} with two possible outcomes: 

\begin{enumerate}
	\item Restoring the technical issue related to the ID-DB.
	\item Tolerating the expired ID.
\end{enumerate} 

If the received ID is adequate (authentic, known, and valid) but the AUTH-DB doesn't have a performance authorization for this drone, then \textbf{Protocol 6} is executed. Three outcomes are possible:
\begin{enumerate}
	\item Restoring the technical issue in the AUTH-DB.
	\item Tolerating the missing performance authorization.
	\item Delayed interdiction.
\end{enumerate} 

When the CUAS finds out that the drone is flying outside the area specified in the authorization certificate in the AUTH-DB, it executes \textbf{Protocol 7}. This protocol has one of the following outcomes:
\begin{enumerate}
	\item Bringing the drone back to the authorized area by the operator.
	\item Tolerating the area violation.
	\item Delayed interdiction. 
\end{enumerate}

Finally, when the CUAS finds out that the operator has exceeded the mission time specified in the authorization certificate, it executes \textbf {Protocol 8}. This protocol has one of the following outcomes:
\begin{enumerate}
	\item Terminating the mission by the operator.
	\item Tolerating time violation
	\item Delayed interdiction. 
\end{enumerate}
	
The states \textit{Tolerate ID Failure} and \textit{Tolerate AUTH Failure} can be reached from different states as described above. To keep the diagram clear, we did not draw the related transitions but used the numbers 1 to 6 to refer to them. Apart from this, these states are equivalent to the states \textit{Valid ID} and \textit{Authorized Mission}, respectively. Nonetheless, if called twice (or another number of times depending on the authority policy), a protocol will generate an interdiction approval to avoid deadlocks. This means that the system does not tolerate a UAS operator who fails to clarify his identity repeatedly.

\section{Clarification Protocols}   
\label{sec:protocols}

The clarification protocols aim at resolving unclear cases towards an accurate classification of the detected drone. In this section, we describe these protocols by first explaining when a protocol is triggered. Then, we summarize the possible outcomes of the protocol using a table and describe its flow using a sequence diagram as far as necessary. For simplicity, we use simple natural language to label the exchanged messages. In the text, the labels are in capital letters for clarity.

\vspace{-7 pt}

\subsection{ {Protocol 1: Clarify Missing ID Transmission}}
\label{sec-p1}
The objective of \textbf{Protocol 1} is to clarify the situation when the CUAS does not receive an ID from the drone. In this case, the CUAS accesses the authorization database (AUTH-DB) to check if there is an operator who is authorized to fly in the respective time and zone. If such an entry exists in the database, the corresponding operator is regarded as the potential operator of the detected drone. The CUAS initiates \textbf{Protocol 1} to clarify whether this potential operator the actual one is, i.e., the one operating the detected drone. First, the CUAS sends a message named NO ID BUT POTENTIAL OPERATOR to the authority. This message includes the ID of the potential operator. The authority sends a CHECK/RESTORE ID TRANSMISSION message to the potential operator or his USS. The authority receives one of four responses or no response from the operator as outlined in Table \ref{tab-protocol-1-cases} (CASE 1 to CASE 5). CASE 6 occurs when the CUAS does not confirm ID restoration, which the operator claims to have done in CASE 5. Next, we describe how each of these cases is treated.
\begin{table}[hbt!]
	\caption{Outcomes of Protocol 1 (Clarify Missing ID Transmission)}
	\vspace{-7 pt}
	\label{tab-protocol-1-cases}
	\centering

	\resizebox{8.5 cm}{!}{
	
	\begin{tabular}{|c|c|c|c|c|}\hline
		
	\textbf{} & \textbf{\thead{Operator \\Response}}  & \textbf{\thead{Risk \\ Assessment}}  & \textbf{\thead{Authority Response\\ to Operator}}  & \textbf{\thead{Authority Response \\to CUAS}} \\ \hline \hline
	
	CASE 1 & \thead{No \\ response} 		& No & NA & \thead{Authorize immediate \\interdiction}\\ \hline
	CASE 2 & \thead{I am \\not flying} 		& No & NA & \thead{Authorize immediate \\interdiction}\\ \hline
	CASE 3 & \thead{I am already\\ transmitting my ID} & Yes & \thead{Order mission \\completion or stop} & \thead{Order mission tolerance or \\authorize timed interdiction}\\ \hline 
	CASE 4 & \thead{I am not able \\ to restore ID} & Yes & \thead{Order mission \\completion or stop} & \thead{Order mission tolerance or \\authorize timed interdiction}\\ \hline 
	CASE 5 & \thead{I restored ID \\ transmission} &  No & NA & \thead{ Verify ID \\restoration} \\ \hline \hline
	CASE 6 & \thead{Unconfirmed \\ID restoration}  & Yes	& \thead{Order mission \\completion or stop} & \thead{Order mission tolerance or\\ authorize timed interdiction}\\ \hline

	\end{tabular}}
\vspace{-7pt}
\end{table}

\begin{figure} [hbt!]
	\centering
	\includegraphics[width=8 cm]{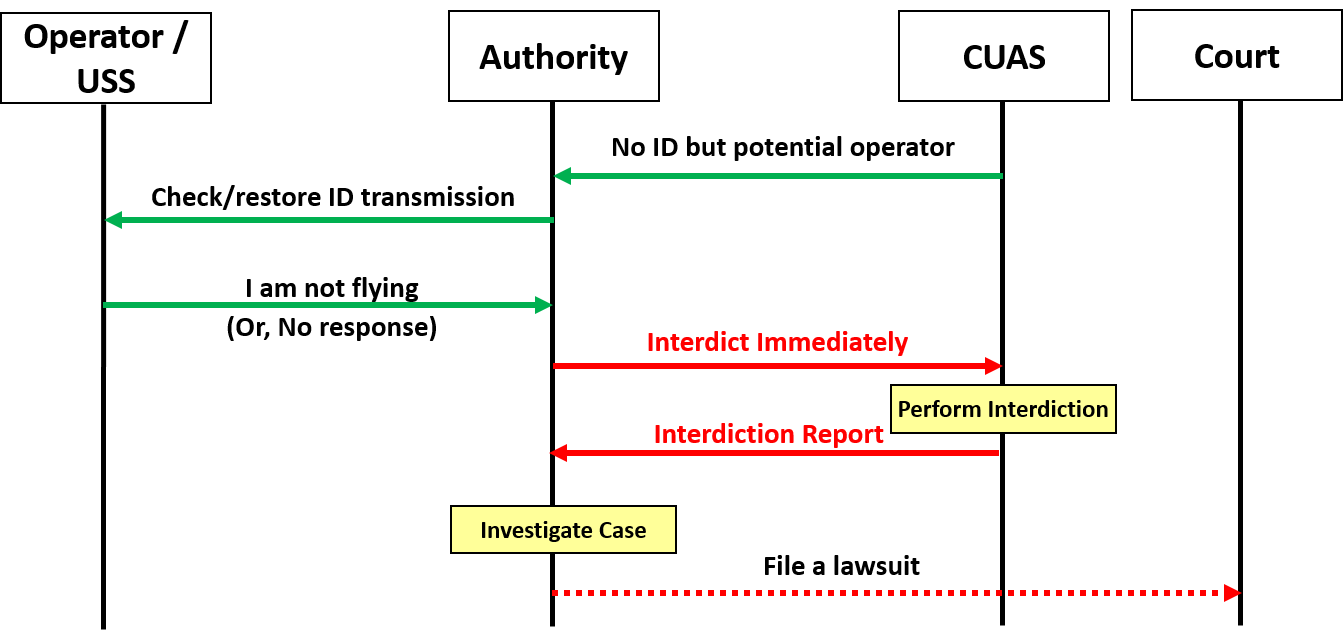}
	\caption{Protocol 1 (Clarify Missing ID Transmission, CASE 1 and CASE 2). Clarify the situation when the CUAS does not receive an ID from the drone. In these two cases, the drone operator either does not respond before the timeout or respond with a "I AM NOT FLYING" message. }
	\label{fig-p1-c1-2}
\end{figure}

In the case of no response (CASE 1) or when the operator confirms that he/she is not flying (CASE 2), the authority may have no possibility for further checks. Hence, it sends an INTERDICT IMMEDIATELY message to the CUAS. The latter performs the interdiction and reports this to the authority, which may file a lawsuit case, see Fig. \ref{fig-p1-c1-2}.

\begin{figure} [hbt!]
	\centering
	\includegraphics[width=8.5 cm]{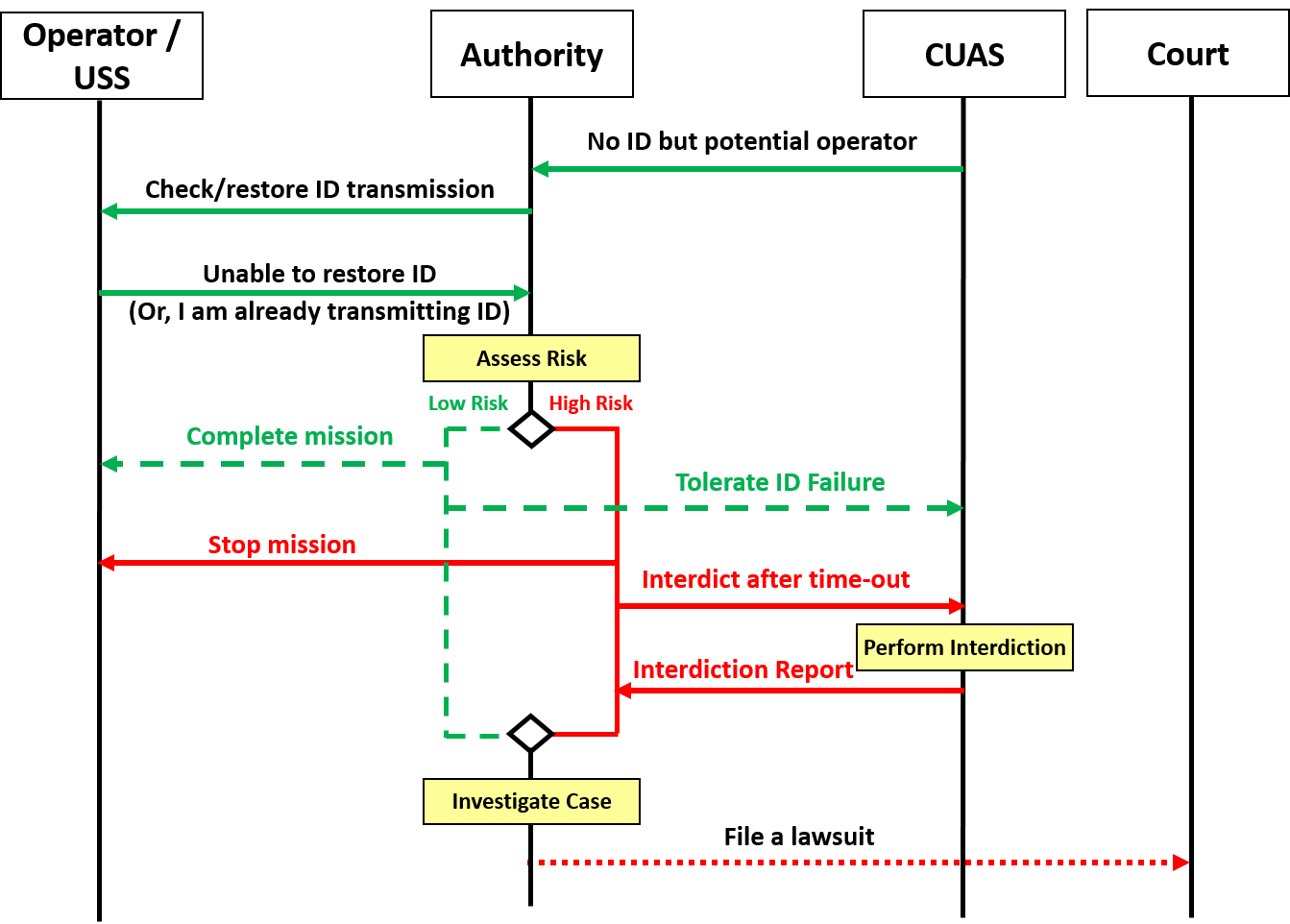}
	\caption{Protocol 1 (Clarify Missing ID Transmission, CASE 3 and CASE 4). The drone operator claims that he/she is unable to restore ID, or that he/she is already broadcasting the ID. }
	\vspace{-10 pt}
	\label{fig-p1-c3-4}
\end{figure}

When the operator confirms that he is already transmitting the ID (CASE 3) or that he is unable to restore his ID transmission for any reason (CASE 4), then the authority must conduct a risk assessment, see Fig. \ref{fig-p1-c3-4}. The risk assessment should take into consideration the criticality of the mission and the no-fly zone. If the estimated risk is low, the authority sends a TOLERATE ID FAILURE message to CUAS and a COMPLETE MISSION message to the operator. In contrast, if the risk is high, the authority sends a STOP MISSION message to the operator and an INTERDICT AFTER TIME-OUT message to the CUAS. Receiving this message, the CUAS can interdict the drone after the time-out specified in this message.  

\begin{figure} [hbt!]
	\centering
	\includegraphics[width=8.5 cm]{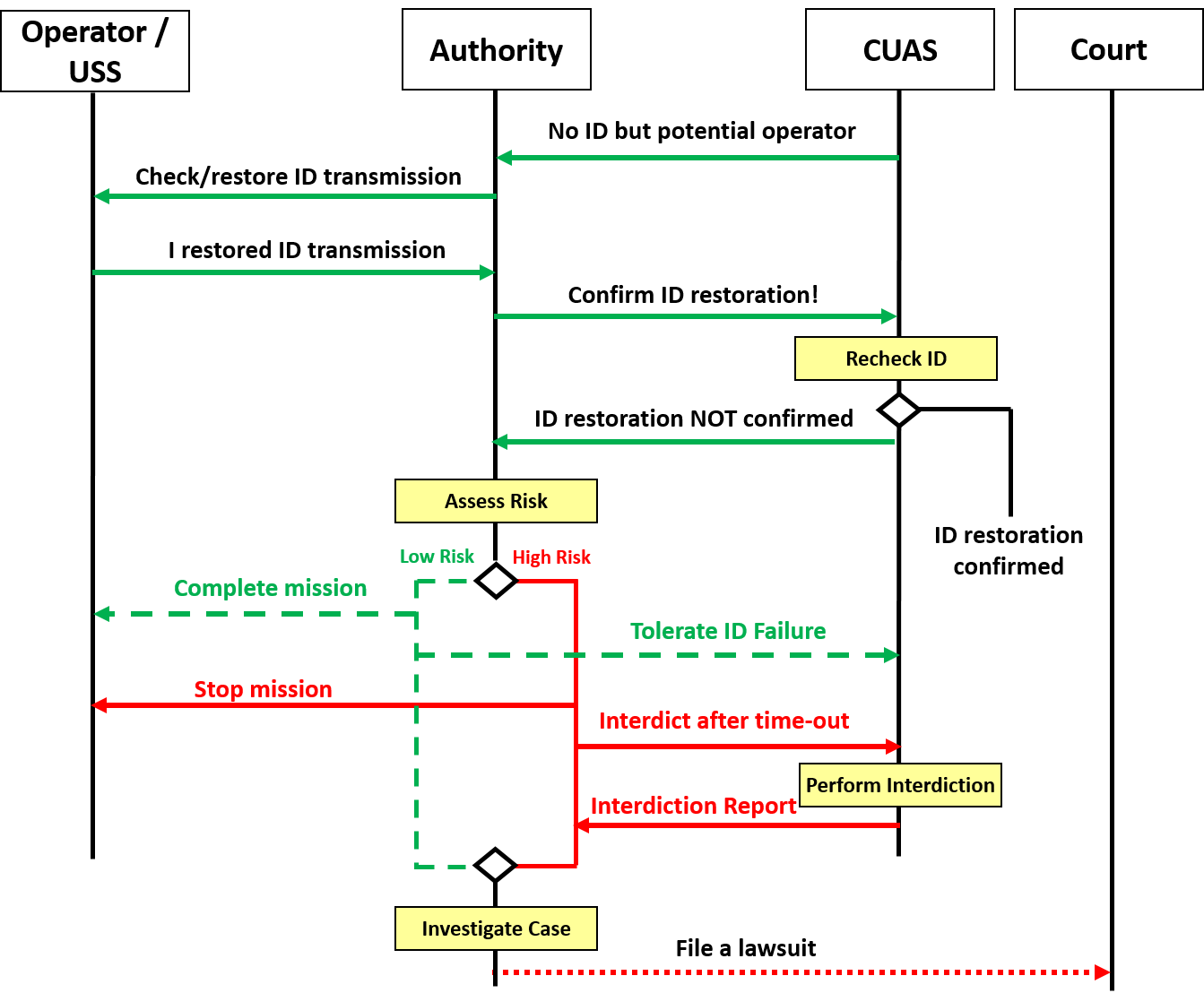}
	\caption{Protocol 1 (Clarify Missing ID Transmission, CASE 5 and CASE 6). The drone operator claims that the ID is restored.}
\label{fig-p1-c5}
\end{figure}

When the operator confirms that he has restored the ID transmission (CASE 5), then the authority asks the CUAS for confirmation by sending a CONFIRM ID RESTORATION! message, see Fig. \ref{fig-p1-c5}. There are two ways to respond to this message: implicit and explicit. In the explicit method, which is illustrated in Fig. \ref{fig-p1-c5}, the CUAS sends dedicated messages (ID RESTORATION CONFIRMED or ID RESTORATION NOT CONFIRMED) to the authority to confirm or invalidate the restoration of the ID transmission, respectively.
Using the implicit method, the CUAS returns to the state \emph{"Drone Detected"} as shown in the post-detection model of Fig. \ref{fig-model}. Now, if the ID transmission was correctly restored, the CUAS would receive the ID and move to the state \emph{"ID Received"}. This would be regarded as an implicit confirmation. In contrast, if the ID restoration was not accomplished, the CUAS would re-send the message NO ID BUT POTENTIAL OPERATOR to the authority. The latter would consider the repeated reception of this message as an invalidation of the ID restoration. An invalidated ID restoration (CASE 6) prompts the authority to perform a fast risk assessment and to act depending on the assessed risk level as illustrated in Fig. \ref{fig-p1-c5}.

\subsection{ {Protocol 2: Clarify Unknown ID Issue}}
\label{sec-p2}
 
The objective of this protocol is to clarify the situation when the CUAS receives an authentic Remote ID but does not find a related entry, neither in the ID-DB nor in the AUTH-DB. This can have technical reasons related to the update or retrieval of these databases. In the worst case, however, receiving an unknown ID can indicate a security break.
First, the CAUS sends an UNKNOWN ID message to the authority. This message includes the received Remote ID. The authority performs necessary checks to find out whether such an ID exists or whether there is a technical issue causing the misses in the databases. Depending on the result of these checks, three outcomes are possible as summarized in Table \ref{tab-protocol-2-cases}.

\vspace{-8pt}
\begin{table}[hbt!]
\caption{Outcomes of Protocol 2 (Clarify Unknown ID Issue)}
\vspace{-8 pt}
	\label{tab-protocol-2-cases}
	\centering
	\resizebox{8.5 cm}{!}{
		
		\begin{tabular}{|c|c|c|c|}\hline
			
			\textbf{} & \textbf{\thead{Description}} & \textbf{\thead{Authority Response\\ to Operator}}  & \textbf{\thead{Authority Response \\to CUAS}} \\ \hline \hline
			
			CASE 1 & \thead{No technical issues, \\ security break} & NA &\thead{Authorize immediate \\interdiction}\\ \hline
			CASE 2 & \thead{Correct ID,\\issue with the ID-BD} & \thead{Order mission \\ stop} & \thead{Authorize timed \\interdiction}\\ \hline
			CASE 3 & \thead{Correct ID,\\issues with both databases} & NA & \thead{Order mission \\ tolerance} \\ \hline

	\end{tabular}}
	\vspace{-10 pt}
\end{table}

In failure-free operation, all authentic IDs are expected to be in the ID database. If the authority finds no technical issues related to the databases, it assumes a security break (CASE 1) because the CUAS has already verified the authenticity of the received ID. For example, the operator can have access to the authority's private key. As a response, the authority sends an IMMEDIATE INTERDICTION AUTHORIZATION message to the CAUS. To keep the drone intact for forensics analysis, the authority may request the CUAS to use a non-destroying technology for the interdiction in this case.

\begin{figure} [hbt!]
	\centering
	\includegraphics[width=8.5 cm]{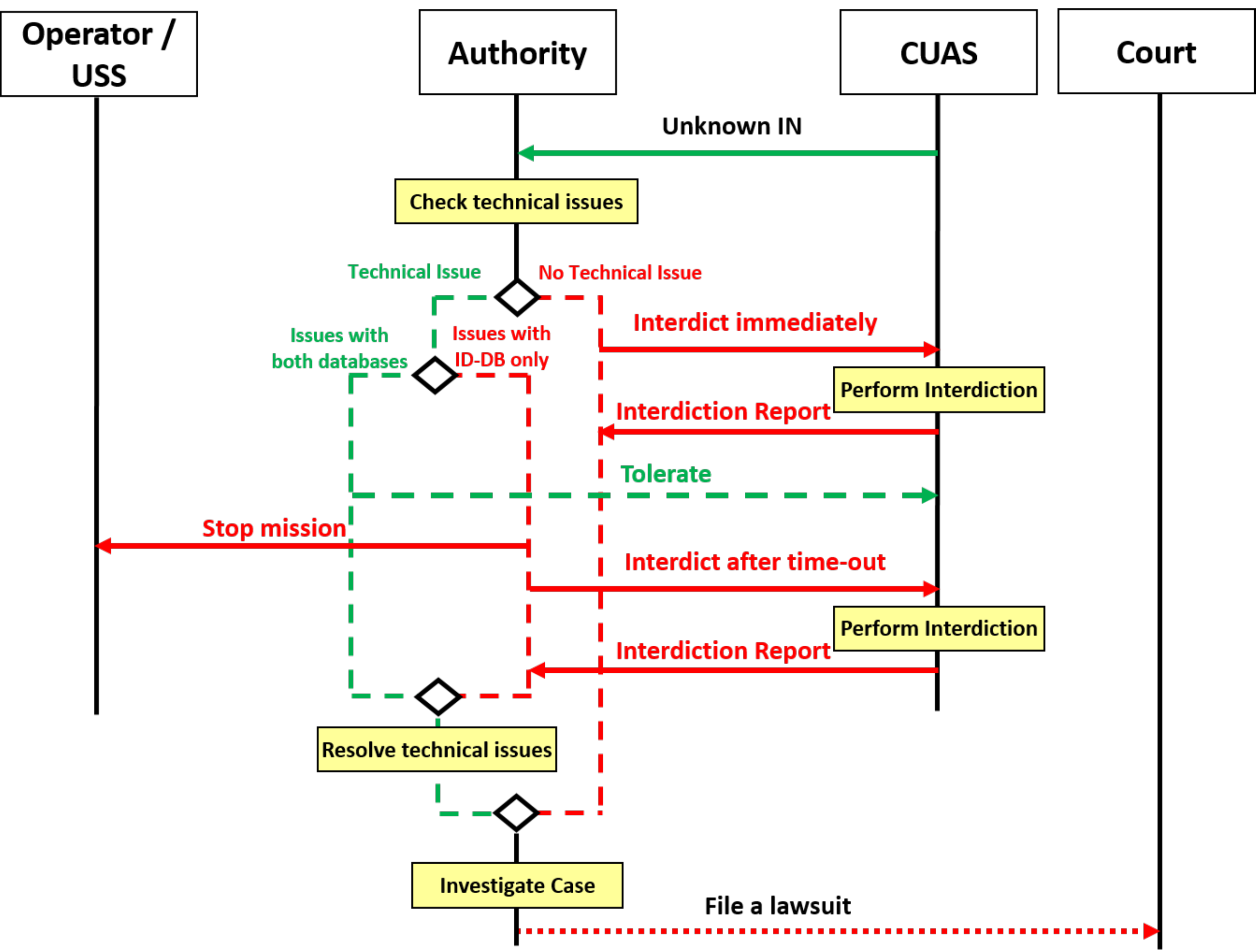}
	\caption{Protocol 2 (Clarify Unknown ID Issue)}
	\label{fig-p2}
	\end{figure}

Alternatively, the authority may find out that the drone is registered but not authorized to fly (CASE 2). This means that the CUAS conclusion regarding the performance authorization was correct but there is a technical failure related to the ID-DB because it did not show the ID. In this case, the authority sends a STOP MISSION message to the operator and an INTERDICT AFTER TIME-OUT message to the CUAS. 

Finally, the authority may identify a technical issue related to the management of both databases (CASE 3). In this case, the CUAS receives a TOLERATE AUTH FAILURE message. This message implies that the ID is authentic, so the CUAS does not need to recheck this.

 \subsection{Protocol 3: Clarify Missing ID in ID-DB}
 \label{sec-p3}
This protocol is used to clarify the situation when the CUAS receives an authentic Remote ID but does not find a related entry in the ID-DB and, at the same time, the AUTH-DB has a performance authorization for this ID in the respective time and area. At first glance, one may attribute this case to an administrative fault. In particular, we may assume that the authority had issued a performance authorization without checking the validity of the ID. While this scenario is possible, it is excluded from our analysis because we assumed in Section \ref{sec-AUTH-DB} that expired IDs are not removed from the database. 

\begin{table}[]
\caption{Outcomes of Protocol 3 (Clarify Missing ID in ID-DB)}
\label{Protocol3}
\centering
\resizebox{7 cm}{!}{
\begin{tabular}{|c|c|c|}
\hline
\textbf{}                    & \textbf{Description}                                                           & \textbf{\begin{tabular}[c]{@{}c@{}}Authority response to CUAS\end{tabular}} \\ \hline \hline 
\multicolumn{1}{|l|}{CASE 1} & \begin{tabular}[c]{@{}c@{}}Unresolved technical \\ issue in ID-DB\end{tabular} & Tolerate ID failure                                                            \\ \hline
CASE 2                       & \begin{tabular}[c]{@{}c@{}}Resolved technical \\ issue in ID-DB\end{tabular}   & Confirm ID restoration                                                         \\ \hline
\end{tabular}}
\vspace{-11 pt}
\end{table}
First, the CAUS sends an ID-BD MISS message to the authority. This message includes the received Remote ID. The authority performs necessary checks. If it identifies a technical issue related to ID-DB management, it sends a TOLERATE ID Failure message to the CUAS and works on resolving the technical issue. Alternatively, the authority can resolve this issue and request the CUAS to confirm this restoration. The latter can respond explicitly or implicitly. Note that the operator is not involved in this protocol. The outcomes of this protocol are summarized in Table \ref{Protocol3}.
\vspace{-10pt}

\subsection{ {Protocol 4: Clarify Unauthorized \& Expired ID}}
\label{sec-p4}

\begin{figure} [hbt!]
	\centering
	\includegraphics[width=8 cm]{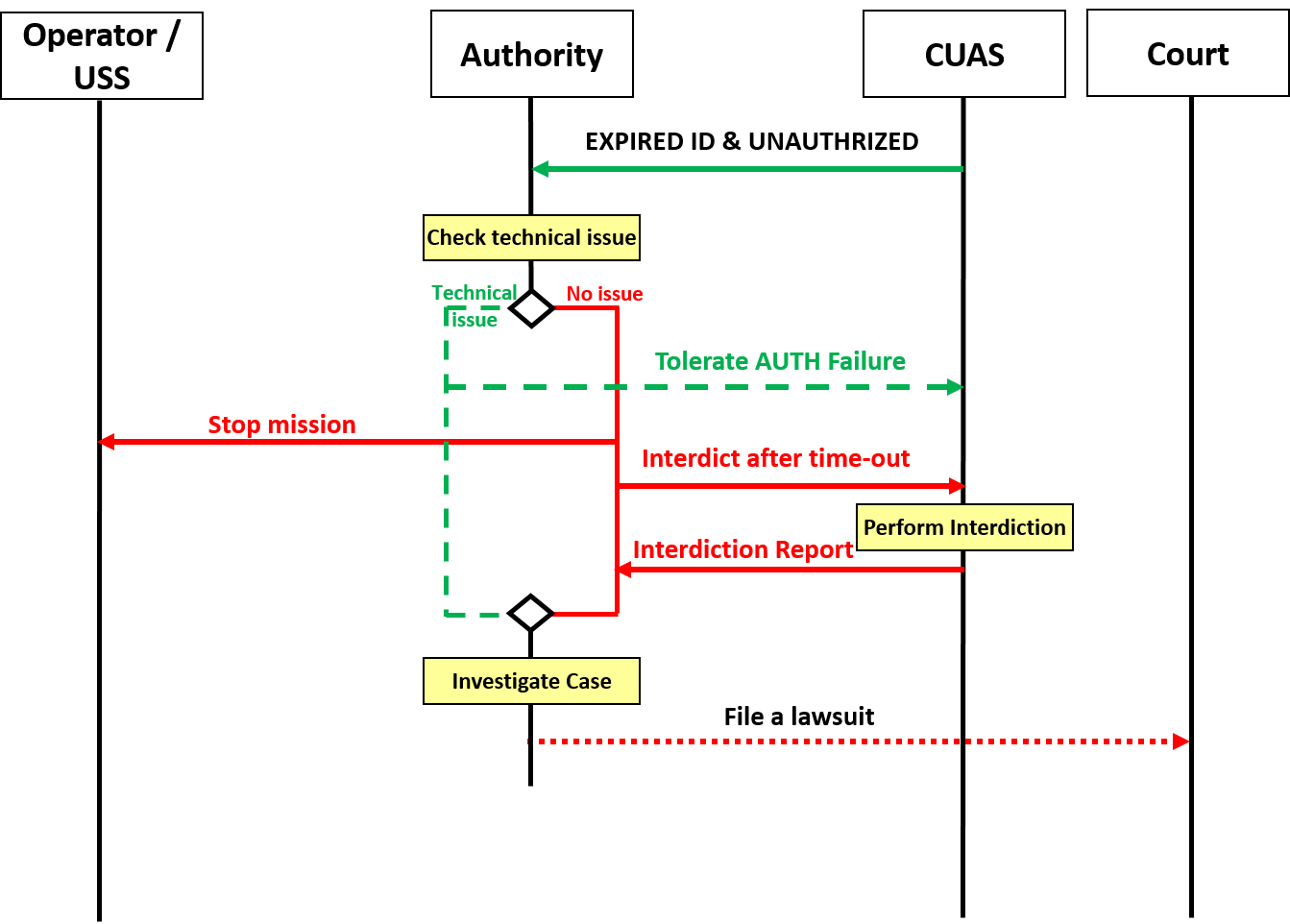}
	\caption{Protocol 4 (Clarify Unauthorized \& Expired ID)}
	\label{fig-p4}
\end{figure}

This protocol deals with a drone that is transmitting an authentic but expired ID. Also, the drone is not authorized to fly. In this case, the CAUS first sends the message EXPIRED ID \& UNAUTHORIZED MISSION to the authority. This message includes the received Remote ID. The authority performs necessary checks to find out if there are any technical issues related to the databases. If no issue is found, the authority sends a STOP MISSION message to the operator and a TIMED INTERDICTION AUTHORIZATION. In contrast, if a technical issue is identified, the authority sends a TOLERATE AUTH FAILURE message to the CUAS and works on fixing the problem.  The message sequence diagram of this protocol is shown in Fig. \ref{fig-p4}.

\subsection{ {Protocol 5: Clarify Authorized But Expired ID}}
\label{sec-p5}

\begin{table}[]
\caption{Outcomes of Protocol 5 (Clarify Authorized But Expired ID) }
\label{Protocol5}
\centering
\resizebox{7 cm}{!}{

\begin{tabular}{|c|c|c|}
\hline
\textbf{}                    & \textbf{Description}                                                           & \textbf{\begin{tabular}[c]{@{}c@{}}Authority response \\ to CUAS\end{tabular}} \\ \hline \hline
\multicolumn{1}{|l|}{CASE 1} & \begin{tabular}[c]{@{}c@{}}Unresolved technical \\ issue in ID-DB\end{tabular} & Tolerate expired ID                                                            \\ \hline
CASE 2                       & \begin{tabular}[c]{@{}c@{}}Resolved technical \\ issue in ID-DB\end{tabular}   & Confirm valid ID entry                                                         \\ \hline

\end{tabular}}
\vspace{-15 pt}
\end{table}

This protocol is executed when the CUAS receives an authentic Remote ID, which is available in the ID-DB but no more valid. However, the AUTH-DB has a performance authorization for this ID. This situation should be attributed to one of two issues: a technical issue in the ID-DB or an administrative mistake by issuing a performance authorization without checking the validity of the ID. 

First, the CAUS sends the message EXPIRED ID BUT AUTHORIZED MISSION to the authority. The latter sends a TOLERATE EXPIRED ID message to the CUAS and works on fixing the problem. Alternatively, the authority updates the ID-DB immediately and requests the CUAS to confirm. The outcomes of this protocol are summarized in Table \ref{Protocol5}.

\subsection{ {Protocol 6: Clarify Missing AUTH}}
\label{sec-p6}
\vspace{-15 pt}
\begin{table}[hbt!]
	\caption{Outcomes of  Protocol 6 (Clarify Missing AUTH)}
	\label{tab-protocol-6-cases}	
	\centering
	\resizebox{8.5 cm}{!}{
		
		\begin{tabular}{|c|c|c|c|c|}\hline
			
			\textbf{} & \textbf{Description}  & \textbf{\thead{Risk \\ Assessment}}  & \textbf{\thead{Authority Response\\ to Operator}}  & \textbf{\thead{Authority Response \\to CUAS}} \\ \hline \hline
			
			CASE 1 & \thead{Technical issue \\ with AUTH-DB} & No & NA & \thead{Technical issue \\resolved}\\ \hline
			CASE 2 & \thead{No technical issue \\ High risk} & Yes & \thead{Order mission \\ stop} & \thead{Authorize timed \\interdiction}\\ \hline 
			
			CASE 3 & \thead{No technical issue \\ Low risk} & Yes & \thead{Order mission \\ completion} & \thead{Order mission \\tolerance}\\ \hline 
			
	\end{tabular}}

\end{table}

This protocol is executed when the received Remote ID is authentic, included in the ID-DB, and valid; but there is no related performance authorization in the AUTH-DB. 

\begin{figure} [hbt!]
	\centering
	\includegraphics[width=8.5 cm]{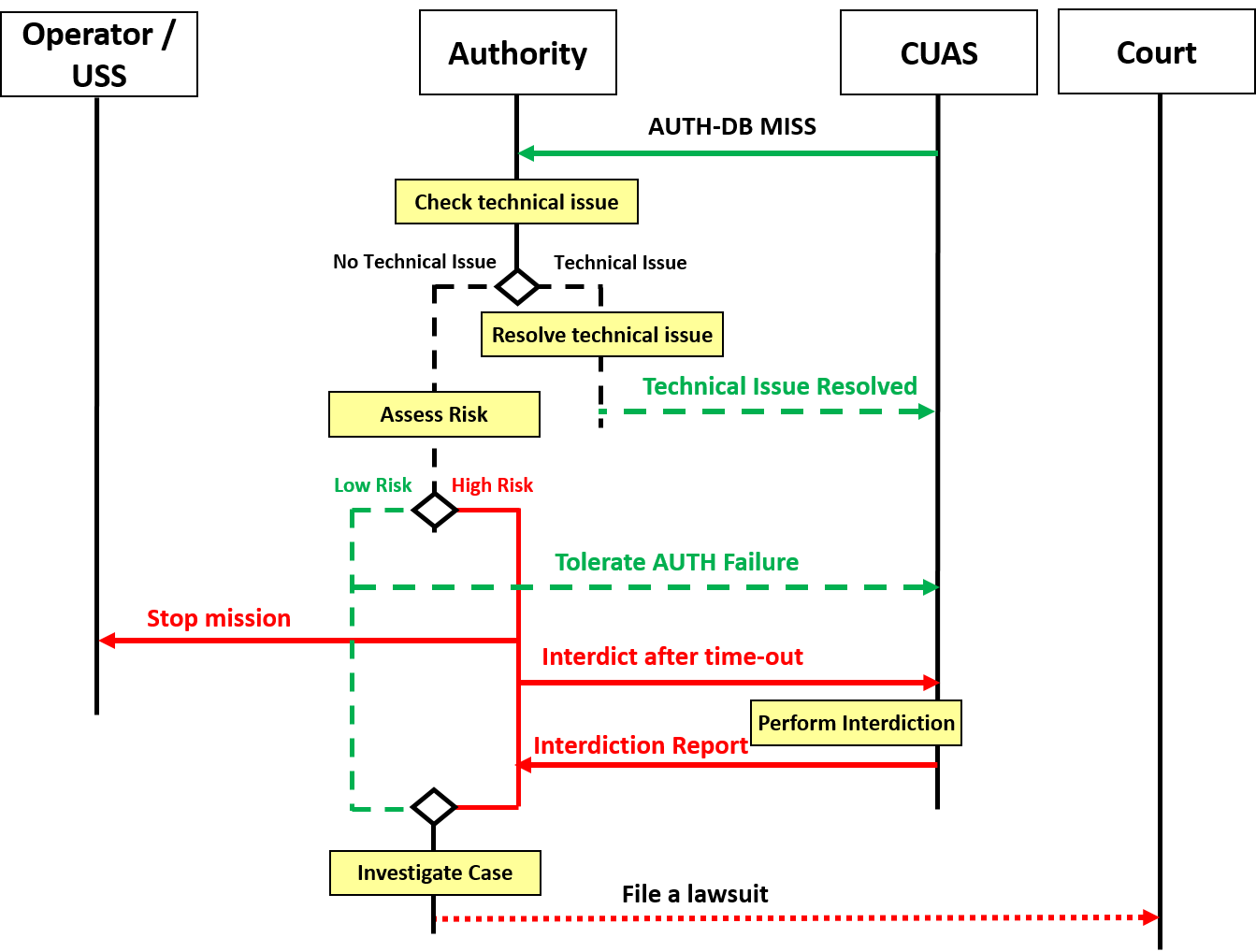}
	\caption{Protocol 6 (Clarify Missing AUTH)}
	\label{fig-p6}
	\vspace{-12 pt}
\end{figure}

To clarify this situation, the CUAS sends an AUTH-DB MISS message to the authority, see Fig. \ref{fig-p6}. The latter checks if the missing authorization is due to a technical problem such as the lack of a timely update (CASE 1), see Table \ref{tab-protocol-6-cases}. If this is the case, the issue is resolved and the CUAS is informed by sending the message AUTH-DB MISS RESOLVED. Otherwise and depending on a risk assessment, the authority requests the operator to stop the mission and the CUAS to interdict after a time-out (CASE 2). Alternatively, the CUAS can be requested to tolerate this mission if the risk is low (CASE 3).

\vspace{-8 pt}
\subsection{ {Protocol 7: Clarify Area Violation }}
\label{sec-p7}
\vspace{-8 pt}
\begin{table}[hbt!]
	\caption{Outcomes of Protocol 7 (Clarify Area Violation)}
	\vspace{-7 pt}
	\label{tab-protocol-7-cases}
	\centering
	\resizebox{8.5 cm}{!}{
		
		\begin{tabular}{|c|c|c|c|c|}\hline
			
			\textbf{} & \textbf{\thead{Operator Response\\ /Description}}  & \textbf{\thead{Risk \\ Assessment}}  & \textbf{\thead{Authority Response\\ to Operator}}  & \textbf{\thead{Authority Response \\to CUAS}} \\ \hline \hline
			
			CASE 1 & \thead{No \\ response} 												& No 	& \thead{Order \\mission stop}	& \thead{Authorize timed \\ interdiction}\\ \hline
			CASE 2 & \thead{I am already flying in \\authorized area} 					& Yes	& \thead{Order mission \\completion or stop} & \thead{Order mission tolerance\\or authorize timed interdiction}\\ \hline
			CASE 3 & \thead{I cannot return \\ to authorized area} 							& Yes 	& \thead{Order mission \\completion or stop} & \thead{Order mission tolerance\\or authorize timed interdiction}\\ \hline
			CASE 4 & \thead{I returned \\ to authorized area} 								& No 	& NA & \thead{Verify return \\ to authorized area} \\ \hline
			CASE 5 & \thead{Unconfirmed return \\to authorized area} 	& No 	& \thead{Order \\mission stop}	& \thead{Authorize timed \\ interdiction}\\ \hline
			
	\end{tabular}}
\end{table}

This protocol is executed when the received Remote ID is adequate and the AUTH-DB has a related performance authorization. However, the operator has flown the drone beyond the authorized zone as specified in the performance authorization. The objective of this protocol is to resolve the issues associated with area violations.
In this case, the CUAS sends an AREA VIOLATION message to the authority. The latter requests the operator to RETURN TO AUTHORIZED AREA. Table \ref{tab-protocol-7-cases} summarizes five possible cases for clarifying this situation.

\begin{figure} [hbt!]
	\centering
	\includegraphics[width=8.5 cm]{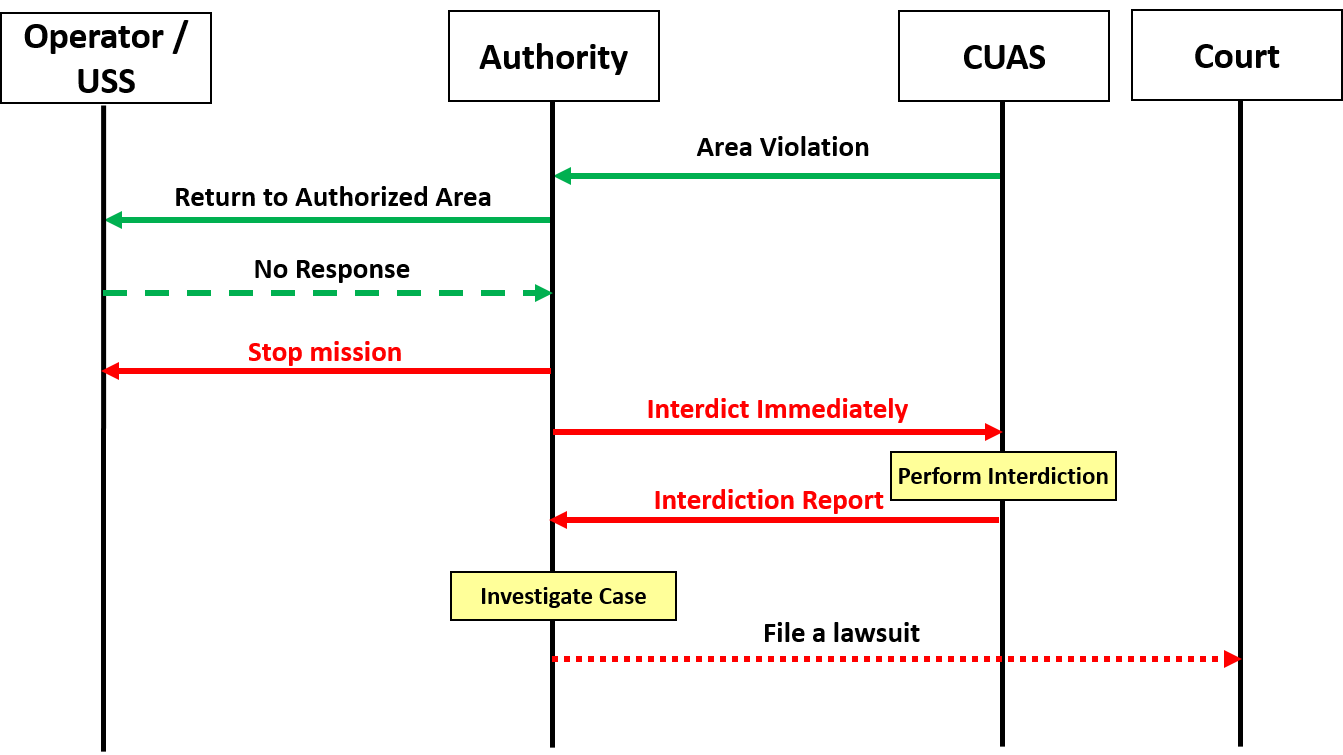}
	\caption{Protocol 7 (Clarify Area Violation, CASE 1). Resolve the issues associated with area violations. This scenario happens when the operator does not respond to the authority message.}
	\label{fig-p7-c1}
\end{figure}

If the operator does not respond to this message (CASE 1), the authority requests the operator to stop the mission and authorizes the CUAS to interdict the drone after a time-out, see Fig. \ref{fig-p7-c1}.

\begin{figure} [hbt!]
	\centering
	\includegraphics[width=8.5 cm]{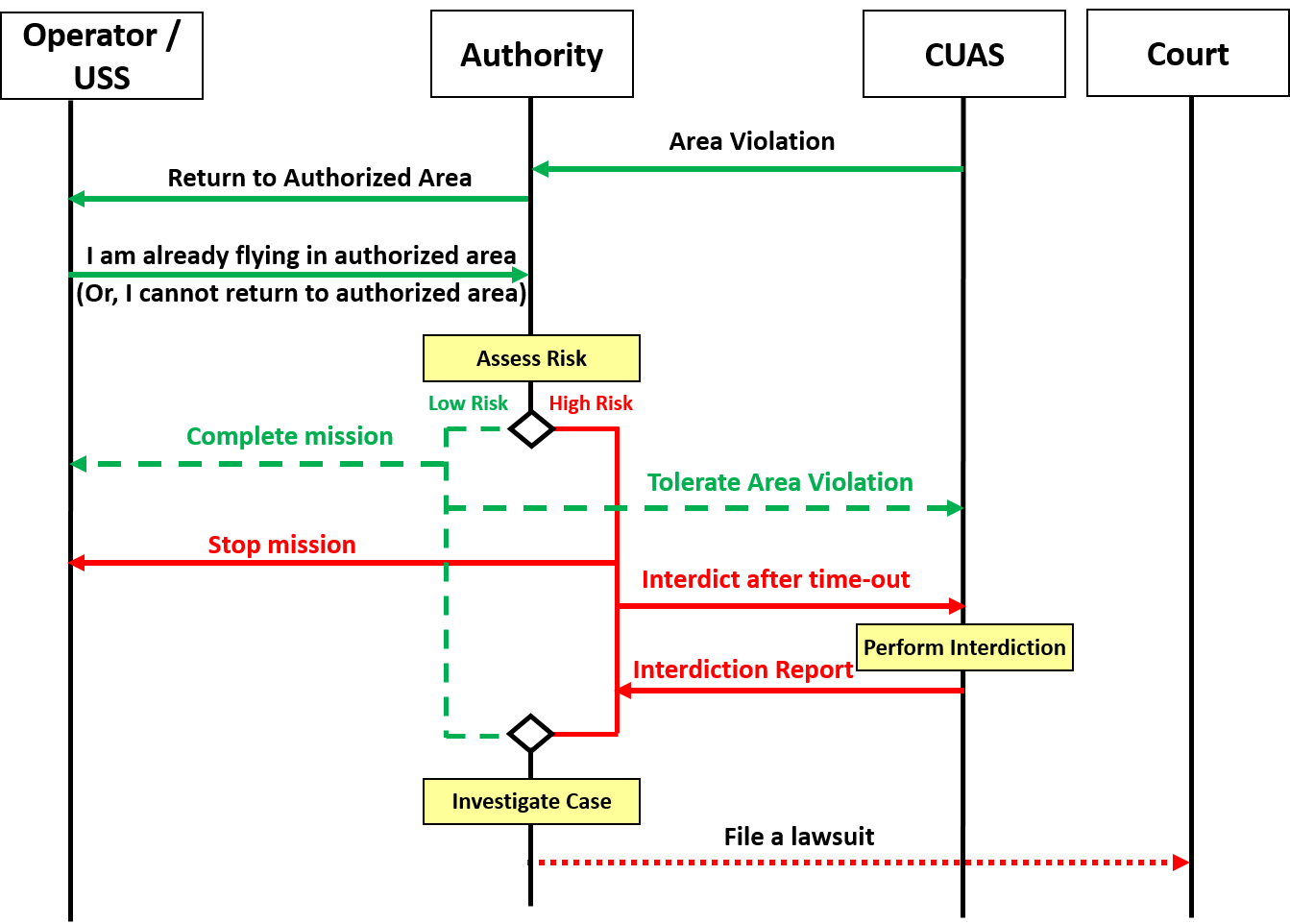}
	\caption{Protocol 7 (Clarify Area Violation, CASE 2 and CASE 3).The operator responds to the authority message.}
	\label{fig-p7-c2-3}
\end{figure}

When the operator contradicts the request by claiming that he did not exceed the permitted area (CASE 2) or when he declares that he cannot return to the authorized zone (CASE3), the authority must perform a fast risk assessment to take an appropriate decision as detailed in Fig. \ref{fig-p7-c2-3}. 

In CASE 4, the operator confirms that he returned to the authorized area. The authority asks the CUAS for confirmation. This confirmation can be explicit or implicit similar to what was explained in \textbf{ {Protocol 1}}. If this proves untrue, or when the operator keeps violating the area restriction (CASE 5), the authority may decide to request the operator to stop the mission and the CUAS to interdict after a time-out. The sequence diagrams for these cases are similar to the previous ones, hence not given here to avoid repetition.

\subsection{Protocol 8: Clarify Time Violation}\footnote{Protocol 8 is very similar to Protocol 7. The operator response may lead to five possible outcomes with similar cases. For space reasons, we omitted the outcomes table and the sequence diagrams.}
\label{sec-p8}

This protocol is executed when the operator has exceeded the flight time specified in the performance authorization according to the data in the AUTH-DB. To resolve the time violation issue, the CUAS sends a TIME VIOLATION message to the authority. The latter requests the operator to STOP MISSION IF AUTHORIZED TIME IS EXCEEDED.

If the operator does not respond to this message (CASE 1), the authority requests the operator to stop the mission and authorizes the CUAS to interdict the drone after a time-out.
When the operator contradicts the request by claiming that he did not exceed permitted time (CASE 2) or when he declares that he cannot stop the mission for any reason (CASE 3), the authority must perform a fast risk assessment to take an appropriate decision.
In CASE 4, the operator confirms that he stopped the mission. The authority asks the CUAS for confirmation. If the latter invalidates the mission stop (CASE 5), the authority decides to request the operator to stop the mission and the CUAS to interdict after a time-out.




\section{System Simulation and Evaluation}
\label{sec-simulation}

\begin{figure} 
	\centering
	\includegraphics[width=0.6\linewidth]{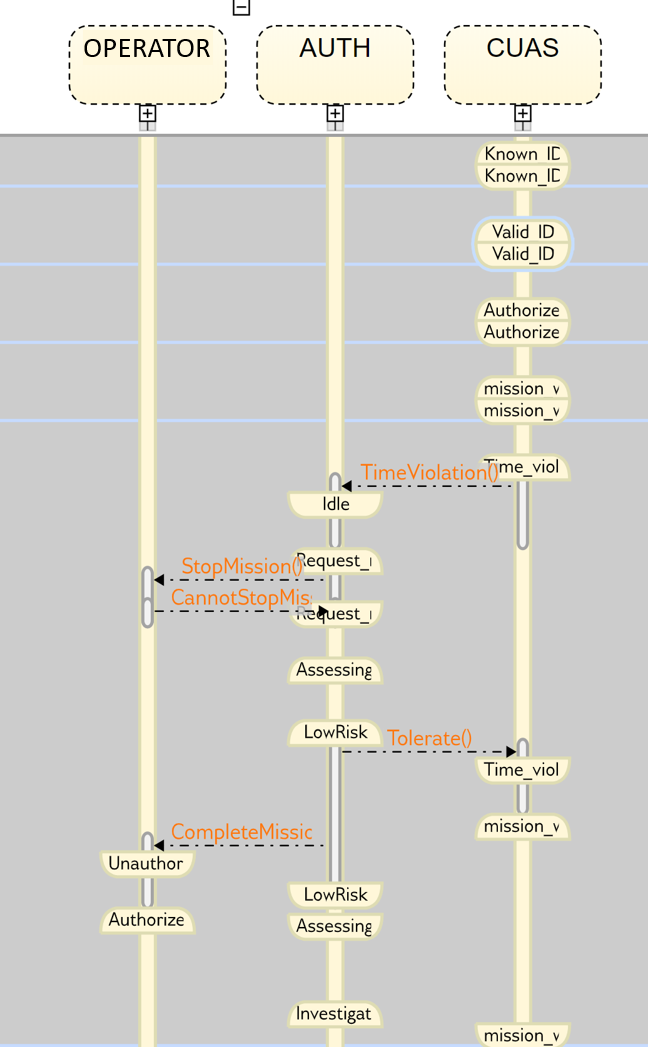}	\vspace{-10 pt}
	\caption{Simulated sequence diagram of Protocol 8 CASE 3  }

	\label{fig. S-V}
\end{figure}

To validate the post-detection model and the clarification protocols, we first developed a high-level model for the system in Matlab. The purpose of this model is to verify the functionality of the proposed system and to identify any errors such as unreachable states, missing transitions, and deadlocks. For this, we used the Stateflow Toolbox \cite{stateflow} that provides a graphical programming language to design and test combinatorial and sequential decision logic using state machines and flowcharts. Besides, this tool offers run-time and edit-time checks to ensure design consistency and completeness before implementation. We created three state machines using this toolbox: the post-detection model that runs on the CUAS side and two additional machines to model the behavior of authority and the operator. The three models were synchronized using global events and executed in parallel to analyze the interaction between the three agents. The toolbox generates sequence diagrams to visualize the system operation as illustrated in the example of Figure \ref{fig. S-V} that shows the sequence diagram of case 3 of Protocol 8.

\begin{figure} 
	\centering
	\includegraphics[width=\linewidth]{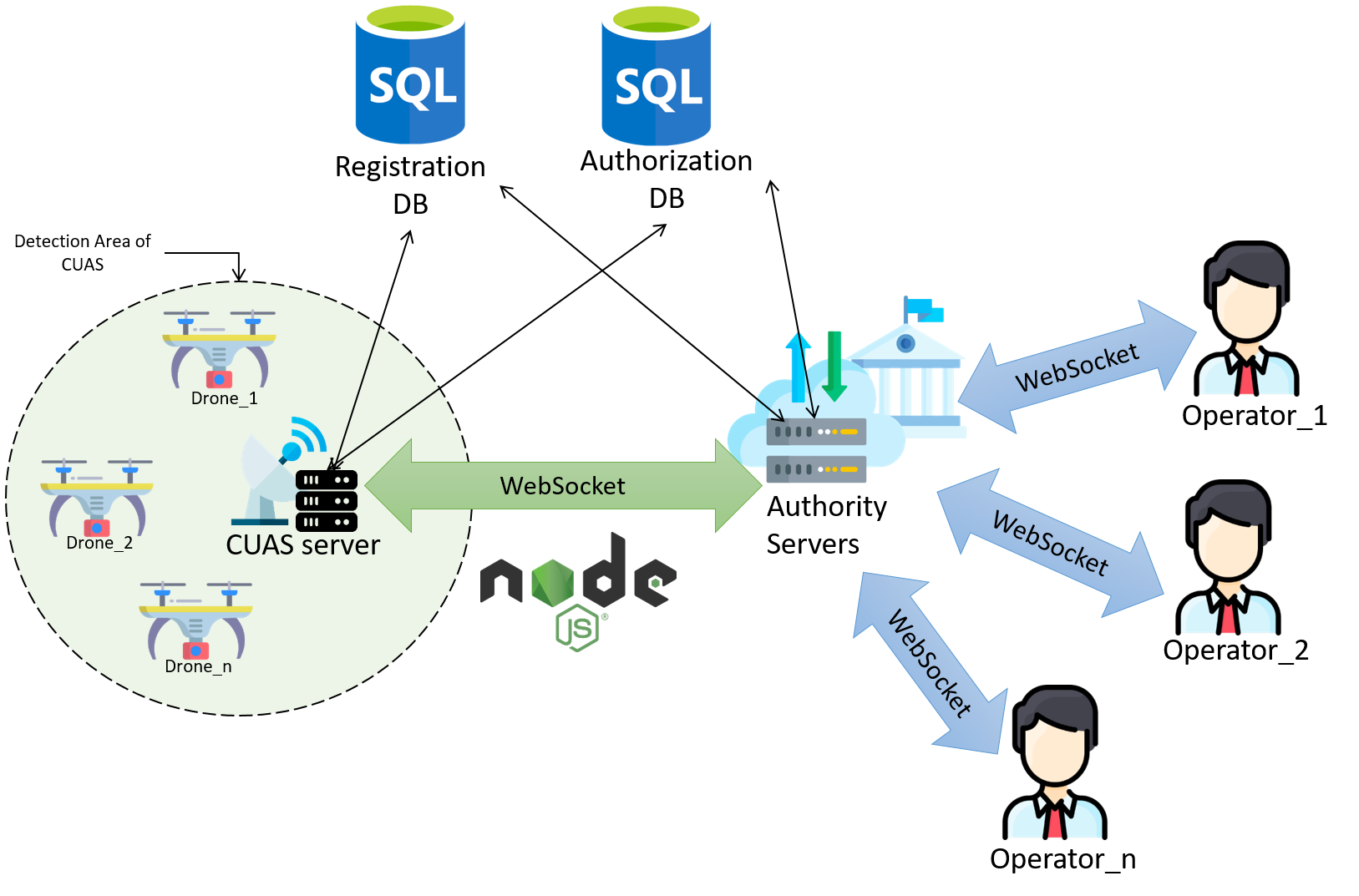}
	\vspace{-20 pt}
	\caption{Overview of the simulation environment.}
	\label{sim}
\end{figure}

\begin{figure*} 
	\centering
	\includegraphics[width=0.8\textwidth]{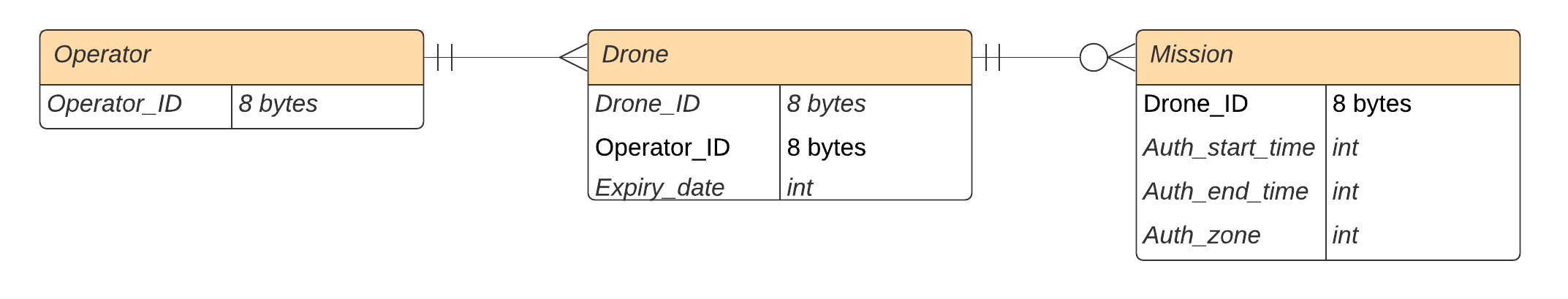}
	\vspace{-10 pt}
	\caption{Entity relationship diagram}
	\label{ERP}
\end{figure*}

In addition to this functional simulation, we implemented the models and protocols and conducted performance testing to evaluate the system behavior in real-time. The goal is to analyze the system performance under different scenarios and with multiple drones. For this purpose, we used Node.js that offers an asynchronous event-driven environment for building scalable applications. In particular, an event loop handles multiple requests simultaneously and invokes a callback function only when a response is returned. We used an open-source library called WebSocket to create a full-duplex connection for message exchange between the server and clients. Since the authority is the central agent in the UTM system, we defined it as the system server. Multiple CUAS and UAV operators can be connected as clients through different ports. The authority and the CUAS operators are granted access to the AUTH-DB and ID-DB databases implemented in SQL. An overview of the simulation setup is illustrated in Figure \ref{sim}. Furthermore, we defined entity relationships between the drone, operator, and missions, as illustrated in Figure \ref{ERP}. This entity-relationship diagram shows that an operator is identified using an 8-byte identifier and can be associated with one or more registered drones. The drone is associated with one or more authorized missions, as long as these missions don't overlap in time. 

We defined simulation scenarios to estimate what we refer to as \emph{clarification time}, which is the time elapsed between detecting a drone by the CUAS system and receiving a related decision from the authority. The simulation scenarios were defined paying attention to complicated cases that account for worst-case clarification times. Multiple scenarios were generated simultaneously to explore the system behavior under peak load conditions. A library called \emph{async} was used in both the CUAS and authority scripts to handle multiple requests in an asynchronous non-blocking mode. We prepared an operator script to model different behaviors throughout the simulation and mimic the different responses to authority messages. Furthermore, we used a Node.js module called \emph{child\_process} to run the operator's script multiple times concurrently. By this means, we could vary the number of detected drones from 1 to 250 in steps of 50 and measure the average clarification time for each protocol. The system was simulated on a local machine with an Intel Core i9-8950HK CPU running at 2.90 GHz and 32-GB RAM.

\begin{figure} [hbt!]
	\centering
	\includegraphics[width=\linewidth]{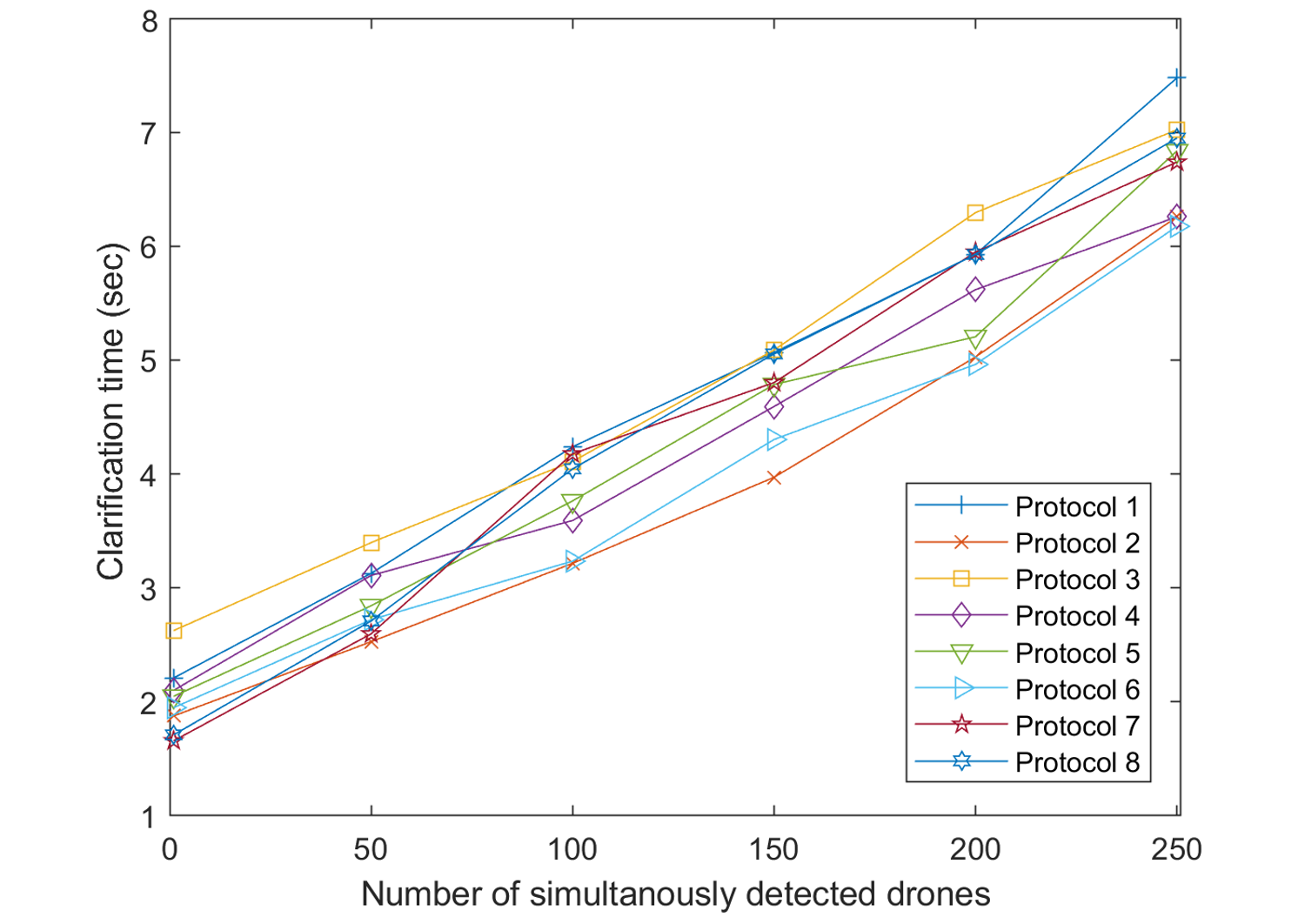}	\vspace{-20 pt}
	\caption{Clarification time of executing the proposed protocols for different numbers of drones.  }

	\label{timedelay}
\end{figure}

\begin{figure} [hbt!]
	\centering
	\includegraphics[width=\linewidth]{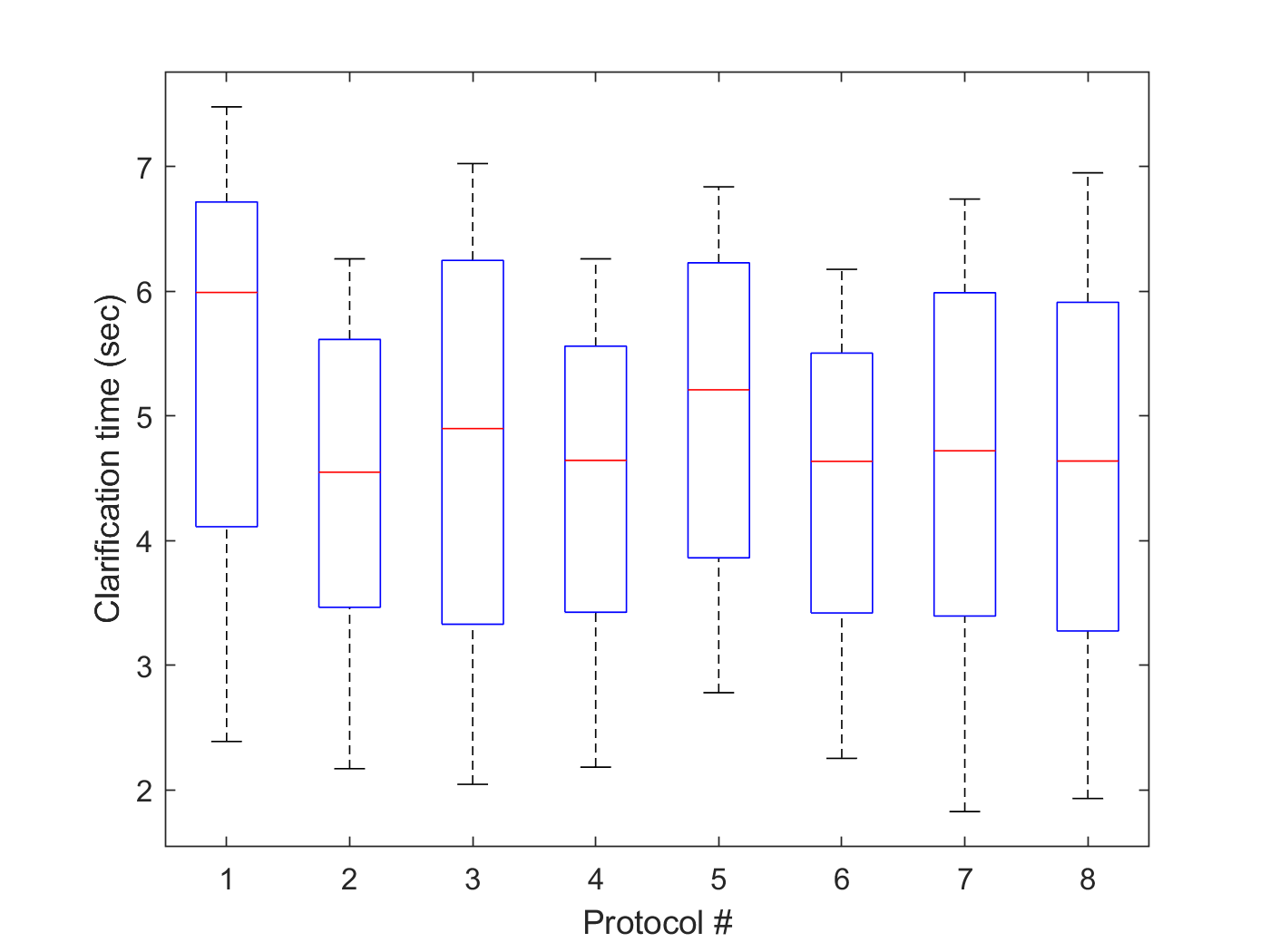}
	\caption{Clarification time of proposed protocols for 250 drones.}
	\vspace{-20pt}
	\label{exec10}
\end{figure}


Figure \ref{timedelay} summarizes the simulation results showing the average clarification time for different protocols and different numbers of drones. Assume, for instance, that the CUAS system has detected 50 drones that all submit their remote identification. However, the CUAS cannot find any of these IDs in the ID-DB. The CUAS would initiate protocol 3 to clarify these cases.  Figure \ref{timedelay} shows that in this case, every drone requires around 3.3 seconds to be clarified, on average. It is important to highlight that this value refers to the average clarification time for \emph{every} drone. So, it does not mean that the 50 drones would require \(3.3\times50 = 165\) seconds. This is due to the system's capability of concurrent processing.
The box-and-whisker plot in Figure \ref{exec10} demonstrates this aspect more clearly. Taking Protocol 1 and the case of 250 drones as an example, we can see that the system can clarify 75\% of the cases in less than seven seconds while the maximum clarification time is 7.5 seconds only. In the next section we discuss how these results should be intepreted. 






\section{Discussion}

To understand the impact of the clarification process on the system performance, we should consider the simulation results in the context of a full counter-drone operation that usually includes three steps: detection/classification, clarification, and interdiction. 
Remember that the system can decide to tolerate the drone as outlined in Figure \ref{fig-model}. In such a case, the clarification time becomes insignificant because there is no need to counter the drone. In contrast, the clarification protocols can affect the system performance when the drone is classified as uncooperative and needs to be interdicted immediately or after a timeout.  So, we have four delay components in total:
\begin{enumerate}
    \item Detection and classification time.
    \item Clarification time
    \item Timeout
    \item Intediction time
\end{enumerate}

We now discuss these time components and try to assign some example values analytically or based on the literature. Reports on the time needed to detect and classify a drone are scarce. Recently, Basak et al. \cite{basak2021combined} proposed a combined drone detection and classification framework using Yolo Lite. They reported a mean inference time of 1.16 seconds for the detection and classification of one drone using a single signal. 
The timeout should be sufficient for at least allowing the operator to land his drone safely. Safe landing takes place at low speeds in the range of a couple of meters per second, e.g., 4 m/sec as reported in \cite{suroso2019analysis}. So, if the drone is at an altitude of 100m, the landing would take around 25 seconds.
The interdiction time depends on the used technology. Jamming is one of the fastest solutions due to its non-kinetic nature. Although the jamming signal should be directed to reduce side effects, the market is rich in off-the-shelf solutions with omnidirectional interceptors that block the communication immediately after switching on. So, the interdiction time of such systems would be negligible. Finally, the simulation results showed that the clarification time is around 2.5 seconds in the case of one drone. 

For these example values, we conclude that the contribution of the clarification time to the total system delay is $2.5/(1.16+2.5)=68\%$, when immediate interdiction is required,  $2.5/(1.16+2.5+25)=9\%$, when the drone should be interdicted after a timeout or $2.5/(1.16+2.5+\infty)=0\%$, when the drone should be tolerated. 

Although this quantitative analysis gives an overall estimation of the proposed solution, it has some limitations which make the results look optimistic or pessimistic. For example, our simulation ran on a single computer. While a real deployment would use distributed systems and servers that are typically faster, network delays would affect the clarification time. Furthermore, we considered a best-case interdiction time of zero which makes our results pessimistic. On the other hand, we did not include the operator's response time during the clarification process. Indeed, real-time measurements on a working implementation of the system with detailed testing scenarios are indispensable for understanding the system's capabilities and challenges. 

Apart from the timing aspects, we concentrated on functional aspects of the system and did not implement security techniques, e.g., for message encryption and authentication. System security is extremely important for this application. Intruders could attack the system and inject fake messages that can lead to drone misclassification with fatal consequences. Finally, in the proposed system,  we did not define the messages' formats. We believe that such activities are beyond the scope of this research and should be addressed collectively by consortiums interested in standardization.

\section {Conclusion}
\label{sec:conclusion}

Currently, developers of UTM systems and counter-drone systems are working in parallel. Drone operators are still benefiting from the current legal situation that prohibits the interdiction of aerial vehicles. Many efforts are on the way to change this situation and we will soon see regulations for counter-drone operations. But like always, regulations are for those who follow them. The system must be prepared to deal with malfunctions, technical faults, and malicious users. The paper showed that the real-time coordination of the drone and counter-drone operation is not a trivial task. The simulation results are promising but real deployments and pilot tests are required to understand the actual dimensions of the problem and the challenges to overcome. We hope that the proposed architecture, models, and protocols will give some directions for system developers to start integrating CUAS and UTM systems and address open issues.




\begin{thebibliography}{10}

\bibitem{faa-final-rule}
Federal~Aviation Administration.
\newblock {UAS Remote Identification Overview}, 2021.

\bibitem{eu1058}
European Union Aviation~Safety Agency.
\newblock {Commission Delegated Regulations (EU) 2020/1058}, 2020.

\bibitem{aker2017using}
Cemal Aker and Sinan Kalkan.
\newblock Using deep networks for drone detection.
\newblock In {\em 2017 14th IEEE International Conference on Advanced Video and
  Signal Based Surveillance (AVSS)}, pages 1--6. IEEE, 2017.

\bibitem{allouch2021utm}
Azza Allouch, Omar Cheikhrouhou, Anis Koub{\^a}a, Khalifa Toumi, Mohamed
  Khalgui, and Tuan Nguyen~Gia.
\newblock Utm-chain: blockchain-based secure unmanned traffic management for
  internet of drones.
\newblock {\em Sensors}, 21(9):3049, 2021.

\bibitem{altawy2016security}
Riham Altawy and Amr~M Youssef.
\newblock Security, privacy, and safety aspects of civilian drones: A survey.
\newblock {\em ACM Transactions on Cyber-Physical Systems}, 1(2):1--25, 2016.

\bibitem{anwar2019machine}
Muhammad~Zohaib Anwar, Zeeshan Kaleem, and Abbas Jamalipour.
\newblock Machine learning inspired sound-based amateur drone detection for
  public safety applications.
\newblock {\em IEEE Transactions on Vehicular Technology}, 68(3):2526--2534,
  2019.

\bibitem{armstrong2019interdiction}
Michael~J Armstrong, George~R Hutchins, and Timothy~A Wachob.
\newblock Interdiction and recovery for small unmanned aircraft systems,
  September~3 2019.
\newblock US Patent 10,401,129.

\bibitem{barrado2020u}
Cristina Barrado, Mario Boyero, Luigi Brucculeri, Giancarlo Ferrara, Andrew
  Hately, Peter Hullah, David Martin-Marrero, Enric Pastor, Anthony~Peter
  Rushton, and Andreas Volkert.
\newblock U-space concept of operations: A key enabler for opening airspace to
  emerging low-altitude operations.
\newblock {\em Aerospace}, 7(3):24, 2020.

\bibitem{basak2021combined}
Sanjoy Basak, Sreeraj Rajendran, Sofie Pollin, and Bart Scheers.
\newblock Combined rf-based drone detection and classification.
\newblock {\em IEEE Transactions on Cognitive Communications and Networking},
  2021.

\bibitem{bekkouche2019toward}
Oussama Bekkouche, Miloud Bagaa, and Tarik Taleb.
\newblock Toward a utm-based service orchestration for uavs in mec-nfv
  environment.
\newblock In {\em 2019 IEEE Global Communications Conference (GLOBECOM)}, pages
  1--6. IEEE, 2019.

\bibitem{chakrabarty2019vehicle}
Anjan Chakrabarty, Corey~A Ippolito, Joshua Baculi, Kalmanje~S Krishnakumar,
  and Sebastian Hening.
\newblock Vehicle to vehicle (v2v) communication for collision avoidance for
  multi-copters flying in utm--tcl4.
\newblock In {\em AIAA Scitech 2019 Forum}, page 0690, 2019.

\bibitem{chin2021efficiency}
Christopher Chin, Karthik Gopalakrishnan, Maxim Egorov, Antony Evans, and Hamsa
  Balakrishnan.
\newblock Efficiency and fairness in unmanned air traffic flow management.
\newblock {\em IEEE Transactions on Intelligent Transportation Systems}, 2021.

\bibitem{courtin2018feasibility}
Christopher Courtin, Michael~J Burton, Alison Yu, Patrick Butler, Parker~D
  Vascik, and R~John Hansman.
\newblock Feasibility study of short takeoff and landing urban air mobility
  vehicles using geometric programming.
\newblock In {\em 2018 Aviation Technology, Integration, and Operations
  Conference}, page 4151, 2018.

\bibitem{Gatwick}
Gordon Darroch.
\newblock Gatwick airport: Drones ground flights.
\newblock {\em BBC}, 2018-20-12.

\bibitem{dedrone-list-of-incidents}
DeDrone.
\newblock Worldwide drone incidents, 2020.

\bibitem{dentler2016real}
Jan Dentler, Somasundar Kannan, Miguel Angel~Olivares Mendez, and Holger Voos.
\newblock A real-time model predictive position control with collision
  avoidance for commercial low-cost quadrotors.
\newblock In {\em 2016 IEEE conference on control applications (CCA)}, pages
  519--525. IEEE, 2016.

\bibitem{drozdowicz201635}
Jedrzej Drozdowicz, Maciej Wielgo, Piotr Samczynski, Krzysztof Kulpa, Jaroslaw
  Krzonkalla, Maj Mordzonek, Marcin Bryl, and Zbigniew Jakielaszek.
\newblock 35 ghz fmcw drone detection system.
\newblock In {\em 2016 17th International Radar Symposium (IRS)}, pages 1--4.
  IEEE, 2016.

\bibitem{Unmanned2019NASA}
FAA.
\newblock Unmanned aircraft system (uas) traffic management (utm) concept of
  operation.
\newblock Technical report, 2020.

\bibitem{laanc}
FAA.
\newblock Uas data exchange (laanc).
\newblock {\em https://www.faa.gov/uas/programs\_partnerships/data\_exchange/},
  2020-29-02.

\bibitem{hassanalian2017classifications}
Mostafa Hassanalian and Abdessattar Abdelkefi.
\newblock Classifications, applications, and design challenges of drones: A
  review.
\newblock {\em Progress in Aerospace Sciences}, 91:99--131, 2017.

\bibitem{kim2017self}
Geon-Hwan Kim, Imtiaz Mahmud, and You-Ze Cho.
\newblock Self-recovery scheme using neighbor information for multi-drone ad
  hoc networks.
\newblock In {\em 2017 23rd Asia-Pacific Conference on Communications (APCC)},
  pages 1--5. IEEE, 2017.

\bibitem{lin2018security}
Chao Lin, Debiao He, Neeraj Kumar, Kim-Kwang~Raymond Choo, Alexey Vinel, and
  Xinyi Huang.
\newblock Security and privacy for the internet of drones: Challenges and
  solutions.
\newblock {\em IEEE Communications Magazine}, 56(1):64--69, 2018.

\bibitem{liu2020iterative}
Huan Liu, Xiamiao Li, Guohua Wu, Mingfeng Fan, Rui Wang, Liang Gao, and Witold
  Pedrycz.
\newblock An iterative two-phase optimization method based on divide and
  conquer framework for integrated scheduling of multiple uavs.
\newblock {\em IEEE Transactions on Intelligent Transportation Systems}, 2020.

\bibitem{lykou2020defending}
Georgia Lykou, Dimitrios Moustakas, and Dimitris Gritzalis.
\newblock Defending airports from uas: A survey on cyber-attacks and
  counter-drone sensing technologies.
\newblock {\em Sensors}, 20(12):3537, 2020.

\bibitem{counter-drone-technologies-survey}
Arthur~Holland Michel.
\newblock Counter-drone systems, 2019.

\bibitem{multerer2017low}
Thomas Multerer, Alexander Ganis, Ulrich Prechtel, Enric Miralles, Askold
  Meusling, Jan Mietzner, Martin Vossiek, Mirko Loghi, and Volker Ziegler.
\newblock Low-cost jamming system against small drones using a 3d mimo radar
  based tracking.
\newblock In {\em 2017 European Radar Conference (EURAD)}, pages 299--302.
  IEEE, 2017.

\bibitem{neogi2021assuring}
Natasha Neogi, Siddhartha Bhattacharyya, Daniel Griessler, Harshitha Kiran, and
  Marco Carvalho.
\newblock Assuring intelligent systems: Contingency management for uas.
\newblock {\em IEEE Transactions on Intelligent Transportation Systems}, 2021.

\bibitem{nguyen2016investigating}
Phuc Nguyen, Mahesh Ravindranatha, Anh Nguyen, Richard Han, and Tam Vu.
\newblock Investigating cost-effective rf-based detection of drones.
\newblock In {\em Proceedings of the 2nd Workshop on Micro Aerial Vehicle
  Networks, Systems, and Applications for Civilian Use}, pages 17--22, 2016.

\bibitem{o2019no}
James O'Malley.
\newblock The no drone zone.
\newblock {\em Engineering \& Technology}, 14(2):34--38, 2019.

\bibitem{park2021survey}
Seongjoon Park, Hyeong~Tae Kim, Sangmin Lee, Hyeontae Joo, and Hwangnam Kim.
\newblock Survey on anti-drone systems: Components, designs, and challenges.
\newblock {\em IEEE Access}, 9:42635--42659, 2021.

\bibitem{reiche2021initial}
Colleen Reiche, Adam~P Cohen, and Chris Fernando.
\newblock An initial assessment of the potential weather barriers of urban air
  mobility.
\newblock {\em IEEE Transactions on Intelligent Transportation Systems}, 2021.

\bibitem{rothe2019concept}
Julian Rothe, Michael Strohmeier, and Sergio Montenegro.
\newblock A concept for catching drones with a net carried by cooperative uavs.
\newblock In {\em 2019 IEEE International Symposium on Safety, Security, and
  Rescue Robotics (SSRR)}, pages 126--132. IEEE, 2019.

\bibitem{ryan2020legal}
Richard Ryan, Saba Al-Rubaye, Graham Braithwaite, and Dimitrios
  Panagiotakopoulos.
\newblock The legal framework of utm for uas.
\newblock In {\em 2020 AIAA/IEEE 39th Digital Avionics Systems Conference
  (DASC)}, pages 1--5. IEEE, 2020.

\bibitem{sandor2019challenges}
Zsolt S{\'a}ndor.
\newblock Challenges caused by the unmanned aerial vehicle in the air traffic
  management.
\newblock {\em Periodica polytechnica transportation engineering},
  47(2):96--105, 2019.

\bibitem{stevens2020geofence}
Mia Stevens and Ella Atkins.
\newblock Geofence definition and deconfliction for uas traffic management.
\newblock {\em IEEE Transactions on Intelligent Transportation Systems}, 2020.

\bibitem{suroso2019analysis}
Indreswari Suroso and Erwhin Irmawan.
\newblock Analysis of uav multicopter of air photography in new yogyakarta
  international airports.
\newblock {\em TELKOMNIKA}, 17(1):521--528, 2019.

\bibitem{taha2019machine}
Bilal Taha and Abdulhadi Shoufan.
\newblock Machine learning-based drone detection and classification:
  State-of-the-art in research.
\newblock {\em IEEE Access}, 7:138669--138682, 2019.

\bibitem{stateflow}
Inc. The~MathWorks.
\newblock Stateflow toolbox, 2021.

\bibitem{wolter2020human}
Cynthia Wolter, Lynne Martin, and Kimberly Jobe.
\newblock Human-system interaction issues and proposed solutions to promote
  successful maturation of the utm system.
\newblock In {\em 2020 AIAA/IEEE 39th Digital Avionics Systems Conference
  (DASC)}, pages 1--7. IEEE, 2020.

\bibitem{wyder2019autonomous}
Philippe~Martin Wyder, Yan-Song Chen, Adrian~J Lasrado, Rafael~J Pelles, Robert
  Kwiatkowski, Edith~OA Comas, Richard Kennedy, Arjun Mangla, Zixi Huang,
  Xiaotian Hu, et~al.
\newblock Autonomous drone hunter operating by deep learning and all-onboard
  computations in gps-denied environments.
\newblock {\em PloS one}, 14(11), 2019.

\bibitem{xu2020recent}
Chenchen Xu, Xiaohan Liao, Junming Tan, Huping Ye, and Haiying Lu.
\newblock Recent research progress of unmanned aerial vehicle regulation
  policies and technologies in urban low altitude.
\newblock {\em IEEE Access}, 8:74175--74194, 2020.

\bibitem{zhou2021control}
Jiazhen Zhou, Dawei Sun, Inseok Hwang, and Dengfeng Sun.
\newblock Control protocol design and analysis for unmanned aircraft system
  traffic management.
\newblock {\em IEEE Transactions on Intelligent Transportation Systems}, 2021.

\end{thebibliography}

\end{document}